\documentclass[aps,pre,reprint,floatfix]{revtex4-1}

\usepackage{amsmath}
\usepackage{amssymb}
\usepackage{algorithm}
\usepackage[noend]{algpseudocode}
\usepackage{graphicx}

\begin{document}

\title{Approximate Bayes learning of stochastic differential
  equations}

\author{Philipp Batz} \email{philipp.batz@tu-berlin.de}
\author{Andreas Ruttor} \email{andreas.ruttor@tu-berlin.de}
\author{Manfred Opper} \email{manfred.opper@tu-berlin.de}
\affiliation{TU Berlin, Fakult\"at IV -- MAR 4-2, Marchstr.~23, 10587
  Berlin, Germany}

\begin{abstract}
  We introduce a nonparametric approach for estimating drift and
  diffusion functions in systems of stochastic differential equations
  from observations of the state vector. Gaussian processes are used
  as flexible models for these functions and estimates are calculated
  directly from dense data sets using Gaussian process regression. We
  also develop an approximate expectation maximization algorithm to
  deal with the unobserved, latent dynamics between sparse
  observations. The posterior over states is approximated by a
  piecewise linearized process of the Ornstein-Uhlenbeck type and the
  maximum a posteriori estimation of the drift is facilitated by a
  sparse Gaussian process approximation.
\end{abstract}

\maketitle

\section{Introduction}

Dynamical systems in the physical world evolve in continuous time and
often the (noisy) dynamics is described naturally in terms of
(stochastic) differential equations \cite{Gardiner:1996:HSM}. However,
due to missing information and/or the complexity of a system it may be
difficult to derive such a model from first principles. Instead, the
goal often is to fit it to observations of the state at discrete
points in time \cite{Iacus:2008:SIS}. So far most inference approaches
for these systems have dealt with the estimation of parameters
contained in the drift function (e.g.~\cite{Wu:2011:BFM} using a
generalized linear model of locally linear forces or
\cite{Golightly:2010:MCM} using a Markov Chain Monte Carlo sampler),
which governs the deterministic part of the microscopic time
evolution. Assumptions for the stochastic part were often simple:
additive noise with the diffusion constant as the only parameter to
estimate. But as both drift and diffusion can be nonlinear functions
of the state vector, a \emph{nonparametric} estimation would be a
natural generalization, when a large number of data points is
available. Previous nonparametric approaches were based on solving the
adjoint Fokker-Planck equation \cite{Lade:2009:FSI} and on kernel
estimators \cite{Bandi:2003:FNE} and are effectively restricted to
one-dimensional models.

An alternative would be a Bayesian nonparametric approach, where prior
knowledge on the unknown functions---such as smoothness, variability,
or periodicity---can be encoded in a probability distribution. A
recent result by \cite{Papaspiliopoulos:2012:NED, Pokern:2013:PCP}
presented an important step in this direction. The authors have shown
that Gaussian processes (GPs) provide a natural family of prior
probability measures over drift functions. If a path of the stochastic
dynamics is observed densely, the posterior process over the drift is
also a GP. Unfortunately, this simplicity is lost, when observations
are not dense, but separated by larger time intervals. In
\cite{Papaspiliopoulos:2012:NED} the case of sparse observations has
been treated by a Monte Carlo approach, which alternates between
sampling complete diffusion paths of the stochastic differential
equation (SDE) and sampling from GP for the drift given a path. A
nontrivial problem is the sampling from SDE paths conditioned on
observations. A second problem stems from the matrix inversions
required by the GP predictions. For a densely sampled hidden path
these matrices become large which leads to a strong increase in
computational complexity. It was shown in
\cite{Papaspiliopoulos:2012:NED} for the case of univariate SDE that
this numerical problem can be circumvented if one chooses a GP prior
where the inverse of the covariance operator is specified as a
differential operator. In this case efficient predictions are possible
in terms of solutions of ordinary differential equations. Recently
\cite{vanderMeulen:2014:RJM} introduced a nonparametric method, which
models the drift as linear combination of variably many basis
functions and uses \emph{reversible-jump Markov chain Monte Carlo} to
sample from the posterior distribution. However, both
\cite{vanderMeulen:2014:RJM} and \cite{Papaspiliopoulos:2012:NED} are
restricted to one-dimensional SDEs. For special cases, where the drift
of the system can be expressed as gradient of a potential,
\cite{Batz:2016:VED} uses the relationship between stationary density
and potential in order to efficiently learn the drift based on the
empirical density of the SDEs.

In this paper, we develop an alternative approximate method for
Bayesian estimation of SDEs based on GPs. The method is faster than
the sampling approach and can be applied to GPs with arbitrary
covariance kernels and also multivariate SDEs. Also, our method is
able to handle non-equilibrium models. In case of dense observations
the framework of GP regression is used to estimate both drift and
diffusion in a nonparametric way. For sparse observations, we use an
approximate expectation maximization (EM) \cite{Dempster:1977:MLI}
algorithm, which extends our approach introduced in the conference
publication \cite{Ruttor:2013:AGP}. The EM algorithm cycles between
the computation of expectations over SDE paths which are approximated
by those of a locally fitted linear model and the computation of the
maximum posterior GP prediction of the drift. In addition, the problem
of the continuum of function values occurring in expectations over the
hidden path is solved by a sparse GP approximation.

The paper is organized as follows. Stochastic differential equations
are introduced in section \ref{sec:sde} and Gaussian processes in
section \ref{sec:gp}. Then section \ref{sec:direct} explains GP based
inference for completely observed paths and shows results on dense
data sets. As large data sets slow down standard GP inference
considerably, section \ref{sec:sparseGP} reviews an efficient sparse
GP method. In section \ref{sec:em} our approximate EM algorithm is
derived and its performance is demonstrated on a variety of SDEs.
Section \ref{sec:discussion} presents a discussion and concludes with
an outline of possible extensions to the method.

\section{Stochastic differential equations and likelihoods for dense
  observations}
\label{sec:sde}

We consider diffusion processes given by a stochastic differential
equation (SDE) written in Ito form as
\begin{equation}
  dX_t = f(X_t) dt + D^{1/2}(X_t) dW_t,
  \label{eq:SDE}
\end{equation}
where the vector function $f(x) = (f^1(x),\ldots, f^d(x))$ defines the
deterministic drift depending on the current state $X_t \in
\mathcal{R}^d$. $W_t$ denotes a Wiener process, which models white
noise, and $D(x)$ is the $d \times d$ diffusion matrix.

Suppose we observe a path $X_{0:T}$ of the process over a time
interval $[0, T]$. Our goal is to estimate the drift function $f(x)$
based on the information contained in $X_{0:T}$. A well known
statistical approach to estimation of unknown model parameters is the
method of maximum likelihood \cite{Iacus:2008:SIS}. This would
maximize the probability of the observed path with respect to $f$. To
derive an expression for such a path probability, we use the Euler
time discretization of the SDE \cite{Kloeden:2011:NSS} given by
\begin{equation}
  X_{t+\Delta t} - X_t = f(X_t) \Delta t + D(X_t)^{1/2} \sqrt{\Delta
    t} \, \epsilon_t,
  \label{eq:dpSDE}
\end{equation}
where $\epsilon_t \sim \mathcal{N}(0, I)$ is a sequence of
i.i.d.~Gaussian noise vectors and $\Delta t$ is a time discretization.
We will later set $\Delta t \to 0$, when we compute explicit results
for estimators. Since the short time transition probabilities of the
process are Gaussian, the probability density for the discretized path
can be written as the product
\begin{equation}
  p(X_{0:T} | f) = p_0(X_{0:T}) L(X_{0:T} | f),
  \label{eq:pathprob}
\end{equation}
where
\begin{equation}
  p_0(X_{0:T}) \propto \exp \left[ -\frac{1}{2 \Delta t} \sum_{t}
    \left|\left| X_{t + \Delta t} - X_t \right|\right|^2 \right]
    \label{eq:Wiener}
\end{equation}
is the measure over paths without drift, and a term
\begin{eqnarray}
  L(X_{0:T} | f) &=& \exp \Bigg[ -\frac{1}{2} \sum_t \left|\left|
      f(X_t) \right| \right|^2 \Delta t \nonumber\\
  &+& \left( f(X_t), X_{t + \Delta t} - X_t \right) \Bigg],
  \label{eq:likelihood}
\end{eqnarray}
which is the relevant term for estimating the function $f$ from the
observations of the path. To avoid cluttered notation, we have
introduced the inner product $(u, v) \doteq u^\top D^{-1} v$ and the
corresponding squared norm $||u||^2 \doteq u^\top D^{-1} u$. The
estimation of $f$ using the method of maximum likelihood can be
motivated by the following heuristics: Consider the case of a very
large observation time $T$. In this limit we may write
\begin{eqnarray}
  &-&\frac{1}{T} \ln L(X_{0:T} | f) \nonumber \\
  &=&\frac{1}{2 T} \sum_t ||f(X_t)||^2 \Delta t - 2
  \left( f(X_t), X_{t + \Delta t} - X_t \right ) \nonumber \\
  &\simeq& \frac{1}{2} \int_0^T \mathrm{E} \left[ ||f(X_t)||^2
  \right] - 2 \mathrm{E} \left[ \left( f(X_t), f_*(X_t) \right)
  \right] dt \nonumber \\
  &=& \frac{1}{2} \int ||f(x)||^2 p(x) dx - \int \left ( f(x),
    f_*(x)\right ) p(x) dx,
  \label{eq:direct}
\end{eqnarray}
where we have taken the limit $\Delta t \to 0$. The expectations are
defined with respect to the true (but unknown) process from which the
data points are generated and $p(x)$ denotes its stationary density.
The true drift is given by the conditional expectation
\begin{equation}
  f_*(x) = \lim_{\Delta t \to 0} \frac{1}{\Delta t} \mathrm{E} \left[
    X_{t + \Delta t} - X_t \middle| X_t = x \right].
    \label{eq:condexpectation}
\end{equation}
Obviously, a minimization of the last term in (\ref{eq:direct}) would
lead to the estimator $\hat{f}(x) = f_*(x)$, which is the true drift
indicating that asymptotically, for a long sequence of data we get a
consistent estimate. Unfortunately, for finite sample time $T$, an
unconstrained maximization of the likelihood (\ref{eq:likelihood})
does not lead to sensible results \cite{Papaspiliopoulos:2012:NED}.
One has to use a regularization approach which restricts the
complexity of the drift function. The simplest possibility is to work
with a parametric model, e.g.~representing $f$ by a polynomial and
estimating its coefficients. However, in many cases it may not be
clear in advance how many parameters such a model should have.

\section{Bayesian estimation with Gaussian processes}
\label{sec:gp}

Another possibility for regularization is a nonparametric Bayesian
approach which uses prior probability distributions $P_0(f)$ over
drift functions. With different choices of the prior different
statistical ensembles of typical drift functions can be selected. We
denote probabilities over the drift $f$ by upper case symbols in order
to avoid confusion with path probabilities. We will also denote
expectations over functions $f$ by the symbol $\mathrm{E}_f$. Our
Bayes estimator will be based on the posterior distribution
\begin{equation}
  p(f | X_{0:T}) \propto P_0(f) L(X_{0:T} | f),
  \label{eqn:GPposterior}
\end{equation}
where the neglected constant of proportionality only contains terms
which do not depend on $f$. To construct such a prior distribution, we
note that the exponent in (\ref{eq:likelihood}) contains the drift $f$
at most quadratically. Hence a natural (conjugate) prior to the drift
for this model is given by a Gaussian measure over functions, i.e.~a
Gaussian process (GP) \cite{Papaspiliopoulos:2012:NED}. Although a
more general model is possible, we will restrict ourselves to the case
where the GP priors over the components $f^j(x)$, $j = 1, \ldots, d$
of the drift factorize and we also assume that we have a diagonal
diffusion matrix $D(x) = \mathrm{diag}(D^1(x), \ldots, D^d(x))$. In
this case, the GP posteriors of $f^j(x)$ also factorize in the
components $j$, and we can estimate drift components independently.

Gaussian processes have become highly popular in Bayesian statistics
especially for applications within the field of machine learning
\cite{Rasmussen:2006:GPM}. Such processes are completely defined by a
mean function $m(x) = \mathrm{E}_f[f(x)]$ (which we will set to zero
throughout the paper) and a kernel function defined as
\begin{equation}
  K(x_1,x_2) = \mathrm{E}_f[f(x_1) f(x_2)],
\end{equation}
which specifies the correlation of function values at two arbitrary
arguments $x_1$ and $x_2$. By the choice of the kernel $K$ we can
encode prior assumptions about typical realizations of such random
functions.

In this paper we will apply Gaussian processes not only to drift
estimation but also to the estimation of the diffusion $D(x)$. The
application in the latter case cannot be entirely justified from a
Bayesian probabilistic perspective, but rather from the point of view
that Gaussian processes are known to provide flexible tools for
nonparametric regression, even when the underlying probabilistic model
is not fully correctly specified. We will give a heuristic derivation
of the analytical results for predictions with Gaussian processes
which is applicable to both drift and diffusion estimation. A more
detailed formulation can be found in \cite{Rasmussen:2006:GPM}. In the
basic regression setting, we assume that we have a set of $n$
input-output data points $(x_i, y_i)$ for $i = 1, \ldots, n$, where
the $y_i$ are modelled as noisy function values $f(X_i)$, i.e.
\begin{equation}
  y_i = f(x_i) + \nu_i,
  \label{eq:GPgeneral}
\end{equation}
where the noise values $\nu_i$, are taken to be independent Gaussian
random variables with zero mean and (possibly different) variances
$\sigma_i^2$. For drift estimation we take $f(x) \equiv f^j(x)$ as an
arbitrary component of the drift vector and setting $D(x) \equiv
D^j(x)$. We then identify
\begin{eqnarray}
  y_i &=& (X_{t_i + \Delta t} - X_{t_i})/\Delta t \\
  \sigma_i^2 &=& \frac{D(x_{t_i})}{\Delta t}.
\end{eqnarray}
In this case, the assumption of Gaussian noise is indeed fulfilled.
Using a GP prior over functions $f$, we try to filter out the noise
from the observations and learn to predict the unobserved function
$f(x)$ at arbitrary input values $x$. For the drift estimation
problem, this equals the conditional expectation
\begin{equation}
  f(x) = \mathrm{E}[X_{t + \Delta t} - X_{t} | X_t = x] / \Delta t
  \label{eqn:GPcondexp}
\end{equation}
for $\Delta t\to 0$. We will discuss the diffusion estimation problem
in the next section, but mention that the noise $\nu_i$ will be no
longer Gaussian. But we will still assume that GP regression will be
able to estimate a conditional expectation of the type
(\ref{eq:GPgeneral}) in this case.

The probabilistic model for regression (\ref{eq:GPgeneral})
corresponds to a likelihood
\begin{equation}
  p(\mathbf{y} | f)  \propto \exp \Bigg[ -\sum_{i=1}^n
  \frac{1}{2\sigma_i^2}  \left(f(x_i) - y_i \right)^2 \Bigg] \ ,
  \label{eq:GPlikelihood}
\end{equation}
It is easy to see, that this likelihood agrees with
(\ref{eq:pathprob}) for the case of drift estimation. To compute the
most likely function $f$ in the Bayesian sense, we minimize the
negative log-posterior functional given by
\begin{eqnarray}
  &-& \ln \left[ P_0(f) p(\mathbf{y} | f) \right] \nonumber \\
  &\simeq & \frac{1}{2} \int \int f(x)K^{-1}(x,x') f(x') dx\; dx'
  \nonumber \\
  && + \sum_{j=1}^n \frac{1}{2 \sigma_i^2}  (f(x_j) - y_j)^2.
  \label{eq:pf}
\end{eqnarray}
Here $K^{-1}$ is the formal inverse of the kernel operator. Setting
the functional derivative
\begin{equation}
  \frac{\delta \ln \left[P_0(f) L(X_{0:T} | f)\right]}{\delta f(x)} =
  0
\end{equation}
and applying the kernel operator $K$ to the resulting equation we get
\begin{equation}
  f(x) = \sum_{j=1}^n \frac{\left(y_j - f(x_j)\right)}{\sigma_j^2}
  K(x,x_j).
\end{equation}
Evaluating this equation at observation $x = x_i$ we obtain
\begin{equation}
  \frac{(y_i - f(x_i))}{\sigma_i^2} = \left( \left( \mathbf{K} +
      \boldsymbol{\Sigma} \right)^{-1} \mathbf{y} \right)_i.
\end{equation}
Here $\mathbf{K} = (K(x_i, x_j))_{i, j = 1}^n$ denotes the kernel
matrix and $\boldsymbol{\Sigma} = \mathrm{diag}(\sigma^2_1, \ldots,
\sigma^2_n)$ is a diagonal matrix composed of the noise variances at
the data points. This yields the following expression (see
\cite{Rasmussen:2006:GPM}) for the GP estimator of the function $f$:
\begin{equation}
  \hat{f}(x) = (\mathbf{k}(x))^\top \left( \mathbf{K} +
    \boldsymbol{\Sigma} \right)^{-1} \mathbf{y},
  \label{eq:predGP}
\end{equation}
where $\mathbf{k}(x) = (K(x, x_i))^\top$. Specializing to the
estimation of the j-th drift component we identify $\mathbf{y} =
((X_{t + \Delta t} - X_t) / \Delta t)^\top$ and $\boldsymbol{\Sigma} =
\mathbf{D}^j/\Delta t$, where $\mathbf{D}^j$ is diagonal matrix
composed of the diffusions $D^j(x_i)$ for $i=1,\ldots,n$, to get
\begin{equation}
  \hat{f}^j(x) = (\mathbf{k}(x)^j)^\top \left( \mathbf{K}^j +
    \frac{1}{\Delta t} \mathbf{D}^j \right)^{-1} \!\!\!\!\!\!
  \mathbf{y}^j,
  \label{eq:dense}
\end{equation}
where $\mathbf{k}(x)^j = (K(x, x_i)^j)^\top$. A similar approach leads
to the Bayesian uncertainty at $x$,
\begin{equation}
  \qquad \hat{D}_{f^j}(x) = K(x, x)^j - (\mathbf{k}(x)^j)^\top \left(
    \mathbf{K}^j + \frac{1}{\Delta t} \mathbf{D}^j \right)^{-1}
  \!\!\!\!\!\! \mathbf{k}(x)^j,
  \label{eq:densediffusion}
\end{equation}
which can be used to quantify the uncertainty of the prediction.

A popular covariance kernel is the radial basis function (RBF) kernel
\begin{equation}
  K_{\tiny{RBF}}(x_1, x_2) = \tau^ 2_\mathrm{RBF} \exp \left(
    -\frac{||x_1 - x_2||^2}{2 l_\mathrm{RBF}^2} \right),
\end{equation}
where the hyperparameters $\tau^ 2_\mathrm{RBF}$ and $l_\mathrm{RBF}$
denote the variance and the correlation length scale of the process.
The RBF kernel assumes smooth, infinitely differentiable functions
$f(\cdot)$. In some cases, the class of functional relationship in the
data set is known beforehand, so that specialized kernel functions
encoding this prior information can be applied. In our experiments, we
use such kernels for the estimation of polynomial and periodic
functions $f(\cdot)$. The corresponding kernels are the polynomial
kernel of degree $p$,
\begin{equation}
  K_{\tiny{Pol}}(x_1, x_2) = \left(1 + x_1^\top x_2 \right)^p,
\end{equation}
and the (one-dimensional) periodic kernel
\begin{equation}
  K(x_1, x_2)_\mathrm{Per} = \tau_\mathrm{Per}^ 2\exp \left( -\frac{2
      \sin \left( \frac{x_1 - x_2}{2}\right)^ 2}{ l_\mathrm{Per}^2}
  \right).
\end{equation}
For the latter kernel, the hyperparameters $\tau^ 2_\mathrm{Per}$ and
$l_\mathrm{Per}$ denote the variance and the correlation length scale
of the process.

In our experiments we found that the choice of the variance kernel
parameter $\tau$ did not have a noticeable impact on the estimation
results. Consequently, we fixed its value to $\tau = 1$. In the case
of the length scale hyperparameter $l$ the user usually has relevant
prior expert knowledge about the specific problem at hand and is able
to determine its value a priori. Similarly, if one knows that the
underlying problem is of polynomial form, one should be able to
specify its order $p$ or at least an upper bound for $p$. We found
that this approach usually works well in practice. We note, however,
that in the case of dense data the kernel hyperparameters can also be
automatically determined in a principled way (see section
\ref{sec:direct}).

\begin{figure}
  \centering
  \includegraphics[width=0.4\textwidth]{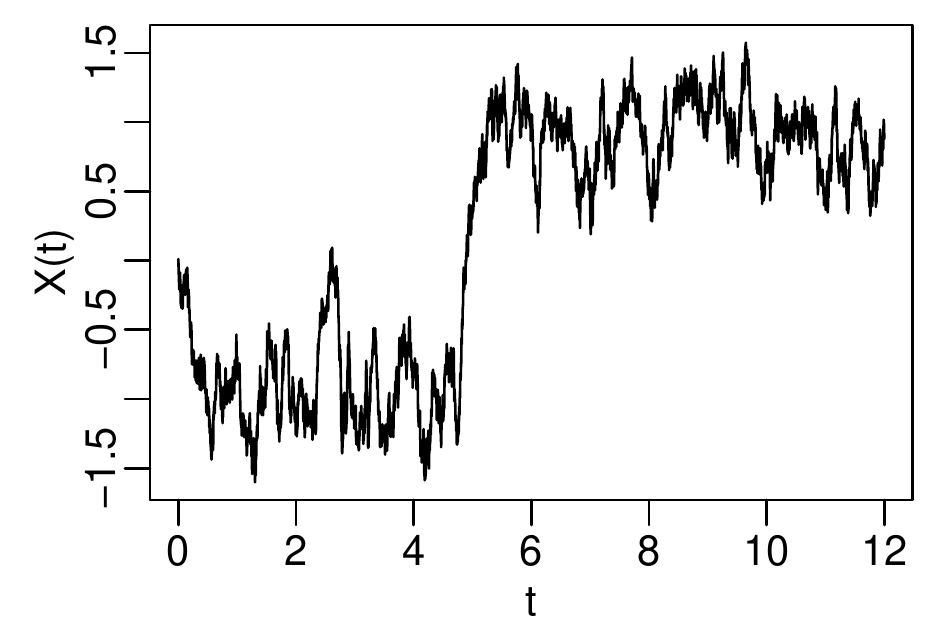}
  \caption{Sample path with $n=6000$ data points generated from a
    double well model with time distance $\Delta_t = 0.002$.}
  \label{fig:DoubleWellFullPath}
\end{figure}

\begin{figure}
  \centering
  \includegraphics[width=0.4\textwidth]{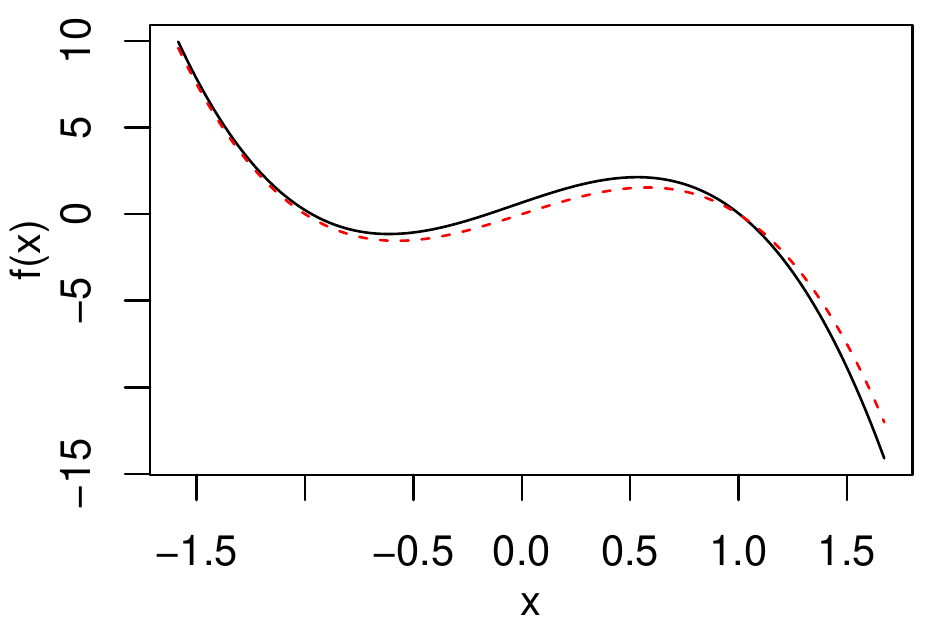}
  \caption{(color online) Estimation for the double well model based
    on the direct GP with the solid black line denoting the mean and
    the dashed red line the true drift function.}
      \label{fig:DWDenseDrift}
\end{figure}

\begin{figure}
  \centering
  \includegraphics[width=0.4\textwidth]{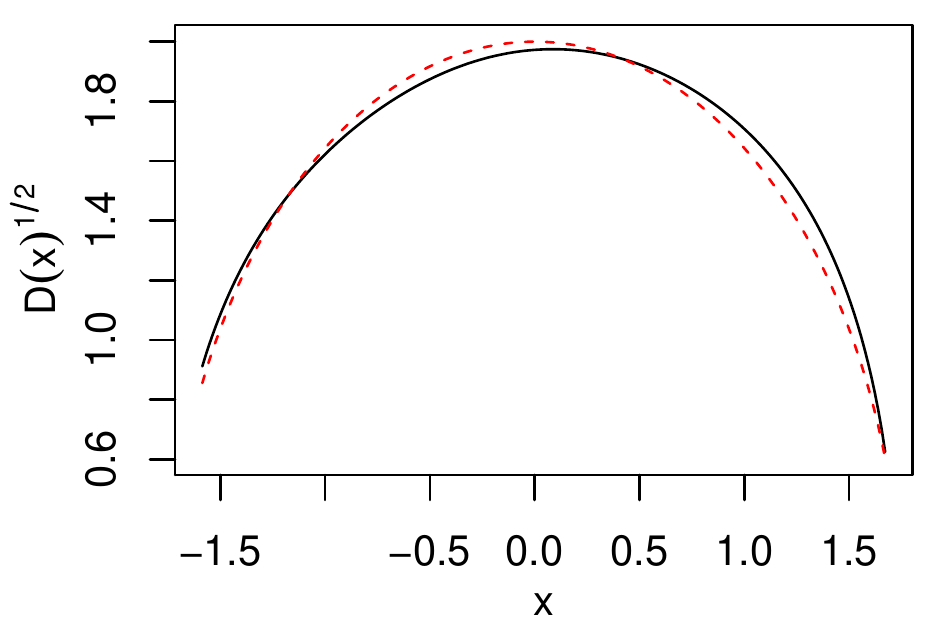}
  \caption{(color online) Diffusion estimation of the double well
    based on the direct GP. The dashed red line denotes the square
    root of the diffusion $D(x)^ {1/2}$ and the solid black line the
    estimator.}
  \label{fig:DWDenseDiffusion}
\end{figure}

\section{Estimation for dense observations}
\label{sec:direct}

Following the dense case of section \ref{sec:gp}, we consider drift
and diffusion estimation in cases where the time grid $\Delta t$ on
which the data points are observed is small. This approach will be
referred to as the \emph{direct Gaussian Process (GP)} estimation with
mean and variance given by (\ref{eq:dense}) and
(\ref{eq:densediffusion}), respectively. We will treat drift and
diffusion estimation in turn and start with the latter. The order is
motivated by the fact that the diffusion estimation is independent of
the drift. Hence, if both drift and diffusion are unknown, one should
first learn the diffusion and then incorporate the estimation results
into the drift learning procedure.

\subsection{Diffusion Estimation}

We distinguish between two cases, namely models with constant and with
state dependent diffusion. If the diffusion matrix $D$ is known to be
constant, i.e.~it does not depend on the state, we will use a Bayesian
maximum likelihood approach and optimize the so-called \emph{Bayes
  evidence}, which equals the probability of the path $p(X_{0:T})$ (in
its Euler discretization), with respect to the diffusion constants $D
=(D^1, \ldots, D^d)$. Again the probability factorizes in the
components $j=1,\ldots,d$. For component $j$ of the process, the
evidence is defined as the $n$-dimensional Gaussian integral
\begin{equation}
  p(X^j_{0:T}) = \int p(X^j_{0:T} | \mathbf{f}^j) p_0(\mathbf{f}^j)
  d\mathbf{f}^j
  \label{eq:evidence}
\end{equation}
where $\mathbf{f}^j$ denotes the vector with components $f^j(X_{t_i})$
for $i=1,\ldots,n$ and $p_0(\mathbf{f}^j) = \mathcal{N}(\mathbf{f}^j
|\mathbf{0},\mathbf{K}^j)$ is the prior Gaussian density induced by
the GP prior over functions. Introducing, as before, the notation
$y^j_i = (X^j_{t_i + \Delta t} - X^j_{t_i})/\Delta t$, we easily
obtain the closed form expression
\begin{equation}
  p(X^j_{0:T}) = \mathcal{N}(\mathbf{y}^j | \mathbf{0}, \mathbf{K}^j +
  \boldsymbol{\Sigma}^j).
\end{equation}
from (\ref{eq:evidence}) with $\boldsymbol{\Sigma}^j = (D^j / \Delta
t) \mathbf{I}$, and where $\mathbf{I}$ denotes the identity matrix.
For the optimization, one can use a standard routine, e.g.~a
quasi-Newton method. The evidence can also be used in the same way to
learn kernel hyperparameters by optimizing with respect to the
specific variables.

In the case of state dependent diffusions $D(x)$, the evidence
optimization becomes impractical, since we would have to jointly
optimize over $D(x_i)$ for all $N$ observations. Instead, we use the
well known representation \cite{Gardiner:1996:HSM} for an arbitrary
component of the exact diffusion
\begin{align}
  D^*(x) &= \lim_{\Delta t \to 0} \frac{1}{\Delta t} \mathrm{Var}
  \left( X_{t + \Delta t} - X_t | X_t = x \right) \nonumber \\
  &= \lim_{\Delta t \to 0} \frac{1}{\Delta t} \left(\mathrm{E} \left[
      (X_{t + \Delta t} - X_t)^ 2 | X_t = x \right]\right. \nonumber
  \\
  &\quad\qquad\qquad\left. - \mathrm{E} \left[ X_{t + \Delta t} - X_t
      | X_t = x \right]^ 2\right) \\
  &= \lim_{\Delta t \to 0} \frac{1}{\Delta t} \left(\mathrm{E} \left[
      (X_{t + \Delta t} - X_t)^ 2 | X_t = x \right]\right. \nonumber
  \\
  &\quad\qquad\qquad \left. -\mathrm{E}[\Delta t f^*(x)]^2 \right)
  \nonumber \\
  &= \lim_{\Delta t \to 0} \frac{1}{\Delta t} \mathrm{E} \left[ (X_{t
      + \Delta t} - X_t)^ 2 | X_t = x \right].
  \label{eq:condvariance}
\end{align}
In the third line, we use the fact that the second term on the right
hand side equals the squared conditional drift
(\ref{eq:condexpectation}). Then---by taking $\Delta t$ out of the
expectation $\mathrm{E}[\Delta t f^*(x)]$---we can easily see that the
term vanishes in the limit $\Delta t \to 0$. Hence, the conditional
variance does not depend on the drift. For its computation, we use
again GP regression, but now on the data set $((x_1, \tilde{y}_1),
\ldots, (x_n, \tilde{y}_n))$, where $\tilde{y}_i=(X_{t_i+\Delta
  t}-X_{t_i})^ 2/\Delta t$ are proportional to the squared
observations of the drift estimation problem.

Unfortunately, the data $\mathbf{\tilde{y}}$ do not follow a Gaussian
distribution, so interpreting the GP posterior as a Bayesian posterior
would lead to a model mismatch. Trying to work with the exact
likelihood would lead to intractable non-Gaussian integrals involving
Gamma-densities which would have to be approximated. Moreover, the
noise in the data $\tilde{y}_i$ depends itself on the diffusion
$D(x_i)$ and a proper Bayesian treatment would lead to a more
complicated iterative estimation problem. However, the observations
$\mathbf{\tilde{y}}^j$ are obviously much smoother than the
$\mathbf{y}^j$. Hence, we expect that the following simpler heuristics
gives good results for densely sampled paths. We regard the GP
framework simply as a regression tool for function estimation, which
in our case happens to be the diffusion function. The regression curve
is given by the GP mean (\ref{eq:predGP}) with $\mathbf{y}^j$
substituted by $\mathbf{\tilde{y}}^j$. Note that this is conceptually
different from the computation of the constant diffusion case above,
where we matched the likelihood variances $\hat{\sigma}_j^2$ to the
diffusion estimators $\hat{D}^j$ of the process. Under the \emph{GP as
  a regression toolbox} lense, the likelihood variance $\sigma^2$
becomes a nuisance parameter without a direct interest to us. Still,
we have to determine suitable variance values as well as possibly
length scale parameters in the case of a RBF kernel, which might not
be readily available.

Finding hyperparameters by optimizing the marginal distribution
presupposes a Bayesian interpretation and so is not applicable in this
context. Therefore we resort to a 2-fold cross-validation scheme. This
method randomly divides the observation into two subsets of equal
size, and learns a GP estimator on each of the subsets. Then the
goodness of fit is determined by computing the mean squared error of
each estimator on the data of the remaining subset.

\subsection{Drift Estimation}

Once we have diffusion values at the observations at our disposal, the
estimation of the drift function becomes straightforward. All we have
to do is to evaluate for each component $j$ the diffusion at the
observations $\mathbf{D}^j(x)$, which we then use as GP variance in
the drift estimation. For the constant but unknown diffusion model, we
insert the estimated value $\hat{D}^j$ into the diagonal of the matrix
$\mathbf{D}^j$, in the state dependent unknown diffusion model, we use
the estimated value $\hat{D}^j(x_i)$ from the diffusion regression
function described above. Then, running the direct GPs on the
observations $\mathbf{y}^j$ leads to a drift estimation, which can
once again be interpreted as Bayesian posterior. However, we emphasize
that we again regard the GP as regression toolbox for computing an
expectation function.

\subsection{Experiments}

Here we show the results for two experiments with unknown state
dependent diffusion. First we look at synthetic data and then at a
real world data set used in climate research. The synthetic data sets
analyzed are generated using the Euler method from the corresponding
SDE with grid size $\Delta t =0.002$.

\subsubsection*{Double well model with unknown state dependent
  diffusion}

In order to evaluate the direct GP method, we generated a sample of
size $n = 5000$ with step size $\Delta t = 0.002$ from the double well
process \cite{Archambeau:2007:VID} with state dependent diffusion:
\begin{equation}
  dX = 4(X - X^3) dt + \sqrt{\max(4 - 1.25 X^2, 0)} dW_t.
\end{equation}
The direct GP was run with a polynomial kernel function of order $p =
4$. The estimation for drift and diffusion function are given in
figures \ref{fig:DWDenseDrift} and \ref{fig:DWDenseDiffusion},
respectively. In both cases, we see a good fit between estimator and
true function.

\subsubsection*{Ice core model}

\begin{figure}
  \centering
  \includegraphics[width=0.4\textwidth]{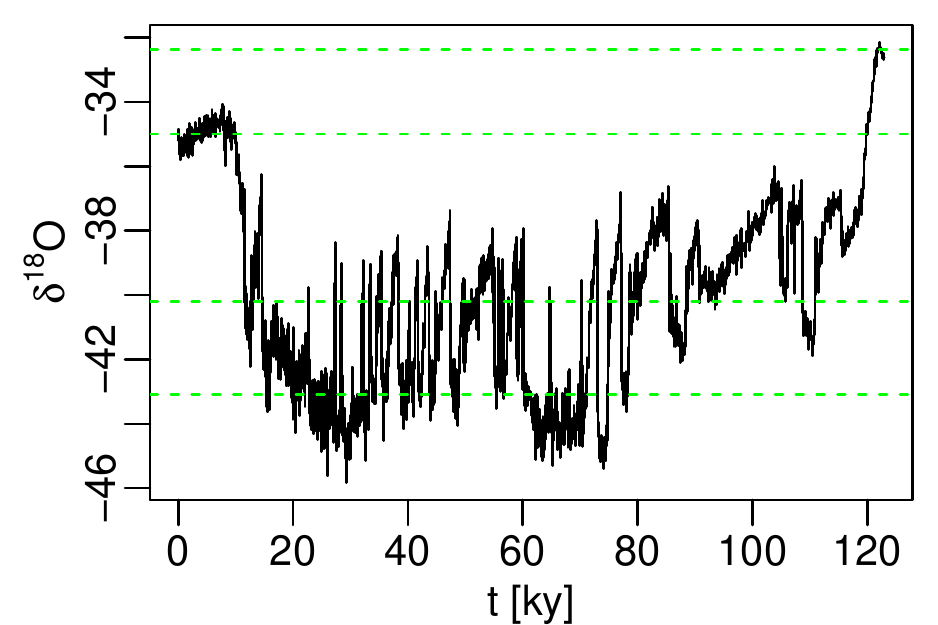}
  \caption{(color online) Plot of the ice core data (as solid black
    line) with metastable states marked by dashed green lines. These
    four minima of the potential function were identified by the
    direct GP algorithm with state dependent diffusion.}
  \label{fig:NGRIP_MetastableStates}
\end{figure}

As an example of a real world data set, we used the NGRIP ice core
data (provided by Niels-Bohr institute in Copenhagen,
\citep{Andersen:2004:HRR}, which provides an undisturbed ice core
record containing climatic information stretching back into the last
glacial. Specifically, this data set as shown in figure
\ref{fig:NGRIP_MetastableStates} contains $4918$ observations of
oxygen isotope concentration $\delta^{18}O$ over a time period from
the present to roughly $1.23 \cdot 10^5$ years into the past. Since
there are generally less isotopes in ice formed under cold conditions,
the isotope concentration can be regarded as an indicator of past
temperatures.

\begin{figure}
  \centering
  \includegraphics[width=0.4\textwidth]{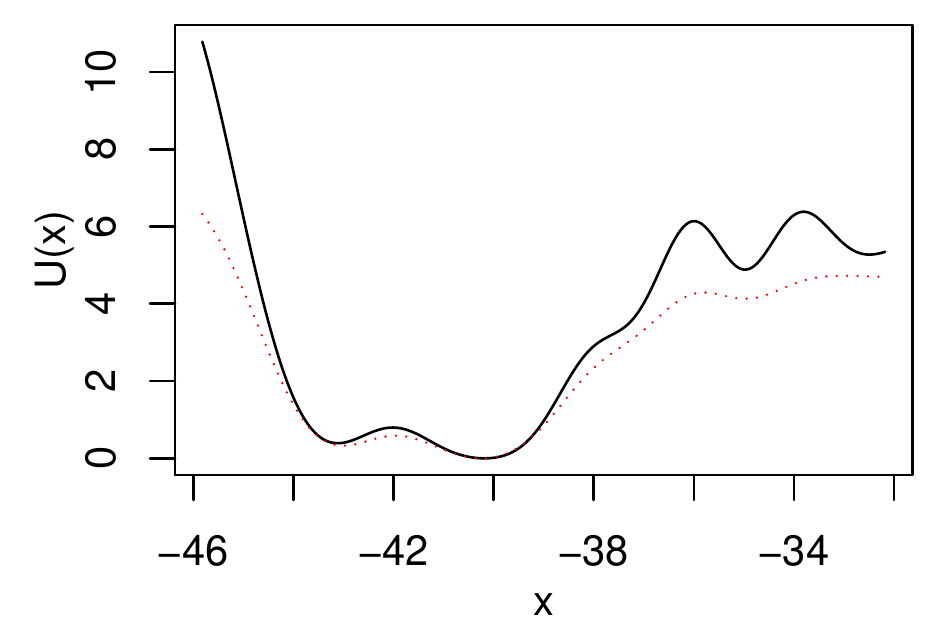}
  \caption{(color online) The figure shows the estimated potentials of
    the ice core data both from a model with state dependent diffusion
    $D(x)$ (solid black line) and with constant diffusion $D$ (dotted
    red). For both models we use a RBF kernel with length scale $l =
    0.7$. The corresponding diffusion estimators are shown in figure
    \ref{fig:NGRIP_Diffusion}.}
  \label{fig:NGRIP_Potential}
\end{figure}

\begin{figure}
  \centering
  \includegraphics[width=0.4\textwidth]{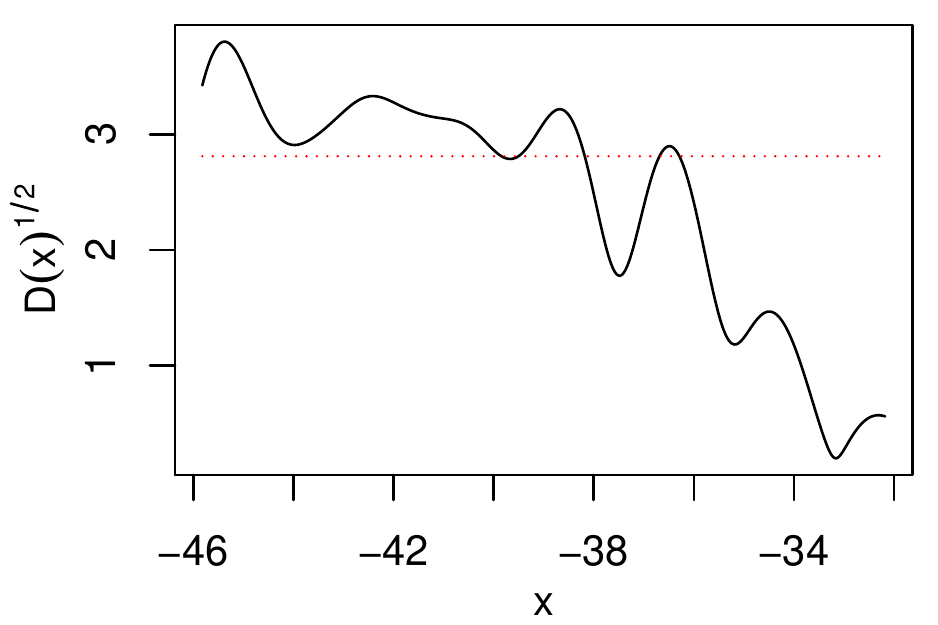}
  \caption{(color online) Diffusion function estimators of the ice
    core model for the state dependent (solid black line) and the
    constant diffusion model (dotted red line). The constant value
    $D^{1/2} = 2.81$ was found by optimization of the marginal
    likelihood. For the GP in the state dependent model we used a RBF
    kernel, whose length scale $l=2.71$ and diffusion $D = 0.1$ was
    determined by 2-fold cross-validation.}
  \label{fig:NGRIP_Diffusion}
\end{figure}

\begin{figure}
  \centering
  \includegraphics[width=0.4\textwidth]{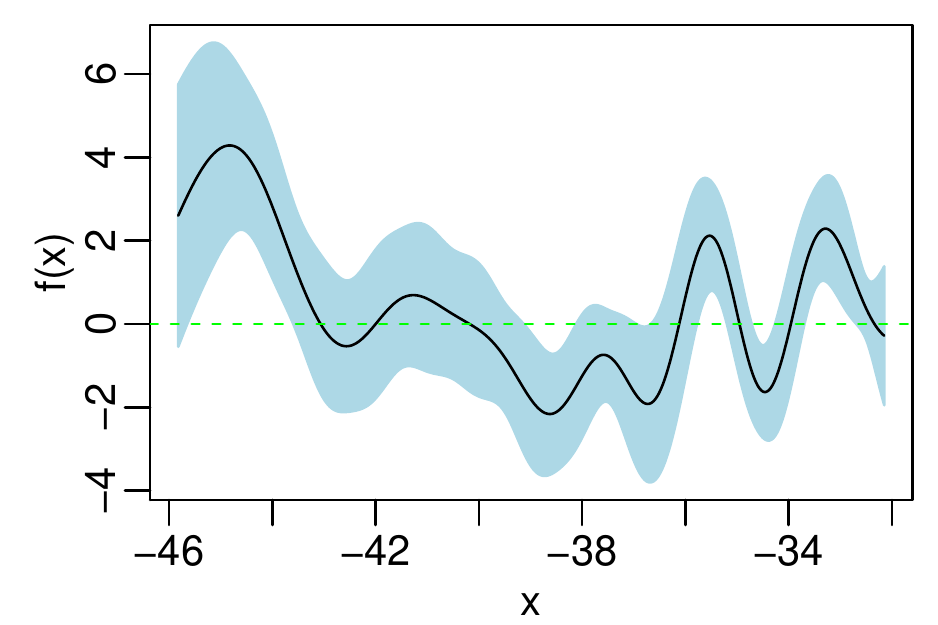}
  \caption{(color online) Plot of the ice core drift function
    corresponding to the above potential function in black together
    with the 95\%- Bayes confidence bounds shaded in blue. One can see
    that the two inner meta stable states are statistically
    significant, while the outer two ones are not.}
  \label{fig:NGRIP_DriftwithConfidence}
\end{figure}

Recent research \cite{Kwasniok:2013:AMG} suggest to model the rapid
paleoclimatic changes exhibited in the data set by a simple dynamical
system with a drift function of order $p = 3$ as canonical model,
which allows for bistability. This corresponds to a meta stable state
at higher temperatures close to marginal stability and a stable state
at low values, which is consistent with other research on this data
set linking a stable state of oxygen isotopes to a baseline
temperature and a region at higher values corresponding to the
occurrence of rapid temperature spikes. For this particular dataset
the consecutive observations are spaced $\Delta t = 0.05
\mathrm{ky}^{-1}$ apart. The underlying dynamics of the NGRIP data set
is often modelled as a constant noise process in the literature
\citep{Kwasniok:2013:AMG}.

Figure \ref{fig:NGRIP_Diffusion} shows that the estimated diffusion
function changes significantly over the range of the observed isotope
concentration, which seems to make the constant diffusion assumption
in the model of \cite{Kwasniok:2013:AMG} inadequate. Our data-driven
approach not only reveals this multiplicative nature of the noise, but
also a richer structure of the learnt potential in comparison to the
potential function of the constant diffusion model, as shown in figure
\ref{fig:NGRIP_Potential}. Hence, choosing a state dependent diffusion
model is advisable even in cases where one is only interested in the
potential, since the noise structure of the data also influences the
estimation of the potential (and drift) function.

Here we find in total four local minima, but only two would be
expected for a polynomial drift of order $p = 3$. Switches between the
two lowest states at $\delta^{18}O \approx -43.1$ and $\delta^{18}O
\approx -40.2$ occur quite frequently due to a low barrier and high
diffusion. A more obvious metastable state is found at $\delta^{18}O
\approx -35.0$ because of an asymmetric barrier and lower noise
levels. As there are only a few data points available around
$\delta^{18}O = -32.4$, this metastable state is not statistically
significant in the estimate of the drift function shown in figure
\ref{fig:NGRIP_DriftwithConfidence}.

\section{Large Number of observations: The need for a sparse GP}
\label{sec:sparseGP}

In practice, the number of observations can be large for a fine time
discretization, and a fast computation of the matrix inverses in
(\ref{eq:dense}) could become infeasible. A possible way out of this
problem---as suggested by \cite{Papaspiliopoulos:2012:NED}---could be
a restriction to kernels for which the inverse kernel is a
differential operator. We will now resort to a different approach
which applies to arbitrary kernels and generalizes easily to
multivariate SDE. Our method is based on a variational approximation
to the GP posterior \cite{Titsias:2009:VLI, Csato:2002:TGF}, where the
likelihood term of the GP model (\ref{eq:likelihood}) is replaced by
another effective likelihood, which depends only on a smaller set of
variables $\mathbf{f}_s$.

\subsection{The general case}

We assume a collection of random variables $f = \{f(x)\}_{x\in T}$
where the index variable $x\in T$ takes values in some possibly
infinite index set $T$. We will assume a \emph{prior measure} denoted
by $P_0(f)$ and a \emph{posterior measure} of the form
\begin{equation}
  P(f) = \frac{1}{Z} P_0(f) \; e^{- U(f)},
\end{equation}
where $U(f)$ is a functional of $f$. The goal is to approximate $P$ by
another measure $Q$ of the form
\begin{equation}
  Q(f) = \frac{1}{Z_s} P_0(f) \; e^{- U_s(\mathbf{f}_s)},
\end{equation}
where the potential $U_s$ depends only on a smaller \emph{sparse} set
$\mathbf{f}_s = \{f(x)\}_{x\in S}$ of dimension $m$. $S$ is not
necessarily a subset of $T$. While we keep the set $S$ fixed, $U_s$
will be optimized to minimize the variational free energy of the
approximation
\begin{equation}
  -\ln Z \leq - \ln Z_s + \mathrm{E}_s\left[U(f) -
    U_s(\mathbf{f}_s)\right].
\end{equation}
We write the joint probability of $\mathbf{f}$ and $\mathbf{f}_s$ as
\begin{equation}
  Q(f,\mathbf{f}_s) = Q(f | \mathbf{f}_s) Q(\mathbf{f}_s) = P_0(f |
  \mathbf{f}_s) Q(\mathbf{f}_s),
\end{equation}
where the last equality follows from the fact that fixing the sparse
set $\mathbf{f}_s$, $U(\mathbf{f}_s)$ becomes non-random and the
dependency on the random variables $f$ is only via $P_0$ and we have
\begin{equation}
  Q(\mathbf{f}_s) = \frac{P_0(\mathbf{f}_s)}{Z_s} \; e^{-
    U_s(\mathbf{f}_s)}.
\end{equation}
Hence, we can integrate out all variables $f$ except $\mathbf{f}_s$
using $P_0(f | \mathbf{f}_s)$ and rewrite the variational bound as the
finite dimensional integral
\begin{eqnarray}
  -\ln Z &\leq& -\ln Z_s + \int Q(\mathbf{f}_s)
  \left\{\mathrm{E}_0[U(f | \mathbf{f}_s)] - U_s(\mathbf{f}_s)\right\}
  d\mathbf{f}_s \nonumber \\
  &=& \int Q(\mathbf{f}_s) \ln
  \left(\frac{Q(\mathbf{f}_s)}{P_0(\mathbf{f}_s) \;
      e^{-\mathrm{E}_0[U(f | \mathbf{f}_s)]}}\right) d\mathbf{f}_s.
\end{eqnarray}
$\mathrm{E}_0[U(f |\mathbf{f}_s)]$ is the conditional expectation
w.r.t.~$P_0$. Since this is of the form of a relative entropy, we
conclude that the bound is minimized by the choice
\begin{equation}
  Q(\mathbf{f}_s) \propto P_0(\mathbf{f}_s) e^{-
    \mathrm{E}_0[U(f|\mathbf{f}_s)]}
  \label{eq:optsparse}
\end{equation}
and thus $U_s(\mathbf{f}_s) = \mathrm{E}_0[U(f |\mathbf{f}_s)]$.

\subsection{Gaussian random variables}

We next specialize to a Gaussian measure $P_0$ with zero mean and
covariance kernel $K$. If we assume (for notational simplicity) that
the set $\{f\}$ is represented as a finite but high-dimensional vector
$\mathbf{f}$ and
\begin{equation}
  U(\mathbf{f}) = \frac{1}{2} \mathbf{f}^\top \mathbf{A} \mathbf{f} -
  \mathbf{b}^\top \mathbf{f}
\end{equation}
is a quadratic form, we can then further simplify the conditional
expectation (\ref{eq:optsparse}) to
\begin{eqnarray}
  \mathrm{E}_0[U(\mathbf{f}) | \mathbf{f}_s] &=& \frac{1}{2}
  (\mathrm{E}_0[\mathbf{f} | \mathbf{f}_s])^\top \mathbf{A}
  \mathrm{E}_0[\mathbf{f} | \mathbf{f}_s] \nonumber \\
  &-& \mathbf{b}^\top \mathrm{E}_0[\mathbf{f} | \mathbf{f}^s] + C,
\end{eqnarray}
where
\begin{equation}
  C = \frac{1}{2} \mathrm{tr} \left(\mathrm{Cov}_0 [\mathbf{f} |
    \mathbf{f}_s] \mathbf{A} \right)
\end{equation}
is a constant independent of $\mathbf{f}_s$. This follows from the
fact that $\mathrm{E}_0[\mathbf{f} | \mathbf{f}_s]$ is the optimal
mean square predictor of the vector $\mathbf{f}$ given $\mathbf{f}_s$
\cite{Papoulis:1965:PRV}, the difference $\mathbf{f} -
\mathrm{E}_0[\mathbf{f}|\mathbf{f}_s]$ is a random vector which is
uncorrelated to the vector $\mathbf{f}_s$ and thus for jointly
Gaussian random variables \emph{independent} of $\mathbf{f}_s$. Hence
the conditional covariance $\mathrm{Cov}_0$ of $\mathbf{f}$ does not
depend on $\mathbf{f}_s$. The explicit result for this predictor is
given by
\begin{equation}
  \mathrm{E}_0[\mathbf{f} | \mathbf{f}^s] = \mathbf{K}_{N s}
  \mathbf{K}_s^{-1} \mathbf{f}_s,
\end{equation}
where $\mathbf{K}_s$ is the kernel matrix for the sparse set and
$\mathbf{K}_{N s}$ is the $n\times m$ kernel matrix between the
non-sparse and the sparse set. It is now easy to generalize to the
infinite dimensional case of the form
\begin{equation}
  U(f) = \frac{1}{2} \int f^2(x) A(x) dx - \int f(x) b(x) dx,
\end{equation}
for which we get
\begin{equation}
  \mathrm{E}_0[f(x) | \mathbf{f}_s] = \mathbf{k}_s^\top(x)
  (\mathbf{K}_s)^{-1} \mathbf{f}_s
\end{equation}
and thus
\begin{eqnarray}
  \mathrm{E}_0[U(\mathbf{f}) | \mathbf{f}_s] &=&
  \frac{1}{2}\mathbf{f}_s^\top \mathbf{K}_s^{-1} \left \{\int
    \mathbf{k}_s(x) \; A(x) \; \mathbf{k}_s^\top(x) dx \right \}
  \mathbf{K}_s^{-1} \mathbf{f}_s \nonumber \\
  &-& \mathbf{f}_s^\top \mathbf{K}_s^{-1} \int \mathbf{k}_s(x) \;
  b(x)\; dx.
\end{eqnarray}

\subsection{Sparse GP Drift and Diffusion Estimation}

Now, setting
\begin{equation}
  U(f) = -\ln[L(X_{0:T}\mid f)],
\end{equation}
we can derive the drift estimator for the sparse representation
analogously to (\ref{eq:pf}). With definitions $\boldsymbol{\pi}^j =
\mathbf{K}_{N s}^j \left(\mathbf{K}_s^j\right)^{-1}$ and $\Omega^j =
\Delta t (\boldsymbol{\pi}^j)^T \mathbf{D}^{-1} \boldsymbol{\pi}^j$ we
get for the $j$th component of the drift vector:
\begin{equation}
  \hat{f}^j(x) = (\mathbf{k}(x)^j)^\top \left(\mathbf{I} + \Omega^ j
    \mathbf{K}_s^j\right)^{-1} \Delta t (\boldsymbol{\pi}^j)^T
  (\mathbf{D}^j)^{-1} \mathbf{y}^j,
  \label{eq:sparsedense}
\end{equation}
where $\mathbf{k}(x)^j = (K(x, x_i)^j)^\top$.

The corresponding expression for the variance estimator is given by:
\begin{equation}
\hat{D}_{f^j}(x) = K(x, x) - \mathbf{k}(x)^\top
  \left(\mathbf{I} + \Omega^j \mathbf{K}_s^j\right)^{-1} \Omega^j
  \mathbf{k}(x).
  \label{eq:sparsedensediffusion}
\end{equation}
Notice that the inverted matrix inside the drift and variance
estimators is no longer of the size of observations $n \times n$, but
of the size of the sparse set $m \times m$.

While it is possible to also optimize the approximation with respect
to the set of sparse points numerically \cite{Csato:2002:SPG,
  Rasmussen:2006:GPM}, we use a simple heuristic, where we construct a
histogram over the observations and select as our sparse set $S$ the
midpoints of all histogram hypercubes containing at least one
observation. Here, the intuition is that a sparse point in a region of
high empirical density is a good approximation to the data points in
the respective hypercube. The number of histogram bins is determined
by Sturges' formula \cite{Sturges:1926:CCI}, which is implicitly based
on the range of the data. Note that the cardinality $m$ of the sparse
set is not set in advance but automatically determined by the spatial
structure of the data. In practice, this heuristic typically leads to
$m\ll n$ and therefore to substantial computational gains compared to
the full GP.

In practice, using the sparse GP for the drift and diffusion function
estimation can be easily accomplished by first determining a sparse
set $S$ for the relevant data set and then substituting mean
(\ref{eq:dense}) and variance (\ref{eq:densediffusion}) equations with
their sparse GP counterparts (\ref{eq:sparsedense}) and
(\ref{eq:sparsedensediffusion}), respectively.

One exception is the estimation of the constant diffusion
$\mathbf{D}$, where we have to replace the marginal distribution
(\ref{eq:evidence}) with a corresponding sparse approximation. Here,
we follow \cite{Titsias:2009:VLI} and optimize for each component $j$
a lower bound to the evidence with respect to the diffusion constants:
\begin{align}
  F_V(X_{0:T}) &=
  \log[\mathcal{N}(\mathbf{y}^j|\mathbf{0},\mathbf{Q}_N^j +
  \frac{1}{\Delta t} \mathbf{D}^j)]\nonumber\\
   &- \frac{\Delta t}{2}
  (\mathbf{D}^j)^{-1}\mathrm{tr}(\mathbf{K}^j-\mathbf{Q}_N^j),
  \label{eq:varievidence}
\end{align}
where $\mathbf{Q}_N^j =
\mathbf{K}_{Ns}^j(\mathbf{K}_s^j)^{-1}(\mathbf{K}_{Ns}^j)^T$ and
$\mathrm{tr}(\cdot)$ denotes the trace of the matrix.

\subsection{Performance comparison}

In order to get a feel for the performance differences between the
standard GP and its sparse counterpart, we compare both versions in
terms of accuracy and performance on the double well model
\begin{equation}
  dX = 4(X - X^3) dt + D^{1/2} dW_t
\end{equation}
with constant and known variance $D = 1$. For the comparison, we
analyzed the performance for data sets of different sizes, where we
generated $10$ data sets with $\Delta t = 0.002$ for each fixed number
of observations. As accuracy measure, we used the approximate mean
squared error (MSE)
\begin{equation}
  \int p(z) (\hat{f}(z) - f(z))^2 dz \approx
  \frac{1}{S} \sum_{i=1}^S (\hat{f}(z_i)
  - f(z_i))^2
\end{equation}
of the corresponding estimator. Here $\hat{f}(z)$ denotes the
estimated drift and $f(z)$ the true drift value, each evaluated on a
set of $S = 100$ fixed points evenly spaced over the range of the
samples. We then measured the run time and MSE of each data set based
on the sparse GP and the standard GP estimation, each with a
polynomial kernel of order $p=4$. All MSE are computed for one fixed
test set of size $n = 4000$, which we generated from the same model
with $\Delta t = 0.5$.

Table \ref{tab:compruntime_mse} shows the mean values of the run time
and MSE for each fixed observation number, respectively. One can see
that the sparse GP algorithm leads to a significant reduction in
computing time while exhibiting practically no loss in estimation
accuracy. As expected, the efficiency gain grows with larger data sets
and even allows us to to analyze big data sets which are
computationally infeasible for the standard GP method.

\begin{table}
  \vspace{1em}
  \centering
  \begin{tabular}{c|cccc}
    Sample & full GP & full GP & sparse GP & sparse GP \\
    Size & Runtime & MSE & Runtime & MSE \\
    \hline
    300 & 0.077 & 1.507 & 0.005 & 1.507 \\
    500 & 0.104 & 1.384 & 0.008 & 1.384 \\
    1000 & 0.828 & 1.292 & 0.014 & 1.293 \\
    2500 & 4.19 & 1.157 & 0.028 & 1.157 \\
    5000 & 30.18 & 0.973 & 0.056 & 0.973 \\
    10000 & 324.5 & 0.592 & 0.162 & 0.593 \\
    50000 & - & - & 0.783 & 0.142 \\
  \end{tabular}
  \caption{Results of mean run times and MSEs of the standard GP and
    sparse GP algorithms for different sample sizes, run on a machine
    with Intel Core i3 processor. The size of the sparse sets varied
    between $m = 6$ and $m = 19$.}
  \label{tab:compruntime_mse}
\end{table}

\section{Estimation for sparse observations}
\label{sec:em}

\begin{figure}
  \centering
  \includegraphics[width=0.4\textwidth]{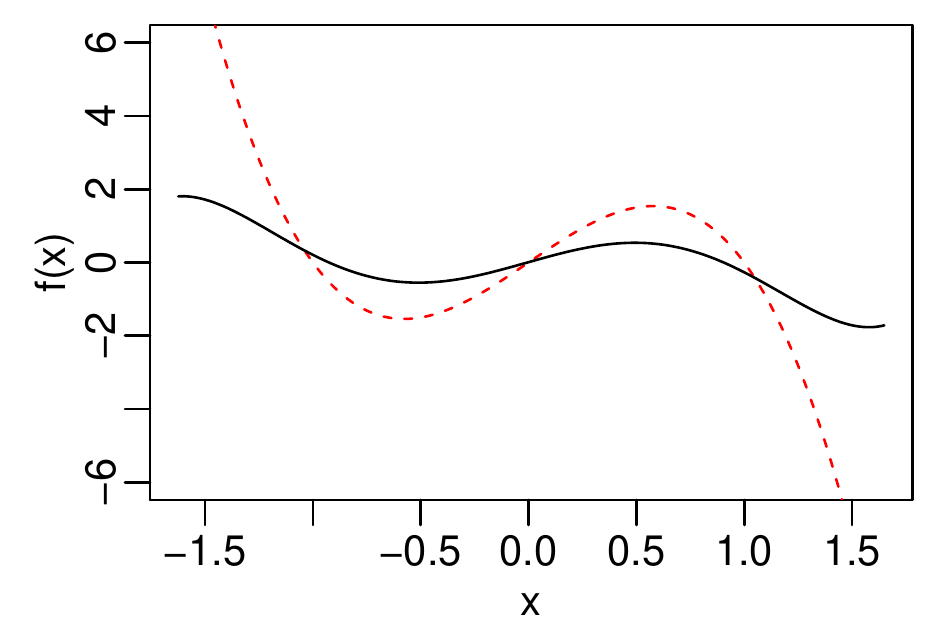}
  \caption{(color online) Estimated drift function for the double well
    based on the direct approach, where the red dashed line denotes
    the true drift function and the solid black line the mean
    function. One can clearly see that the larger distance between the
    consecutive points leads to a wrong prediction.}
  \label{fig:WrongDrift}
\end{figure}

The direct GP approach outlined above leads to wrong estimates of the
drift when observations are sparse in time. In the sparse setting, we
assume that $n$ observations $z_k \doteq X_{\tau_k}$, $k=1,\ldots,n$
are obtained at (for simplicity) regular intervals $\tau_k = k \tau$,
where $\tau \gg \Delta t$ is much larger than the microscopic time
scale. In this case, a straightforward discretization in
(\ref{eq:likelihood}), where the sum over microscopic times $t_i$
would be replaced by a sum over \emph{macroscopic} times $\tau_k$ and
$\Delta t$ by $\tau$, would correspond to a discrete time dynamical
model of the form (\ref{eq:SDE}) again replacing $\Delta t$ by $\tau$.
But this discretization is a bad approximation to the true SDE
dynamics. This is because the transition kernel over macroscopic times
$\tau$ is simply not a Gaussian for a general $f$ as was assumed in
(\ref{eq:pf}). The failure of the direct estimator for larger time
distances can be seen in figure \ref{fig:WrongDrift}, where the red
line corresponds to the true drift of the double-well (with constant,
known diffusion) and the black line to its prediction based on
observations with $\tau = 0.2$.

\begin{figure}
  \centering
  \includegraphics[width=0.4\textwidth]{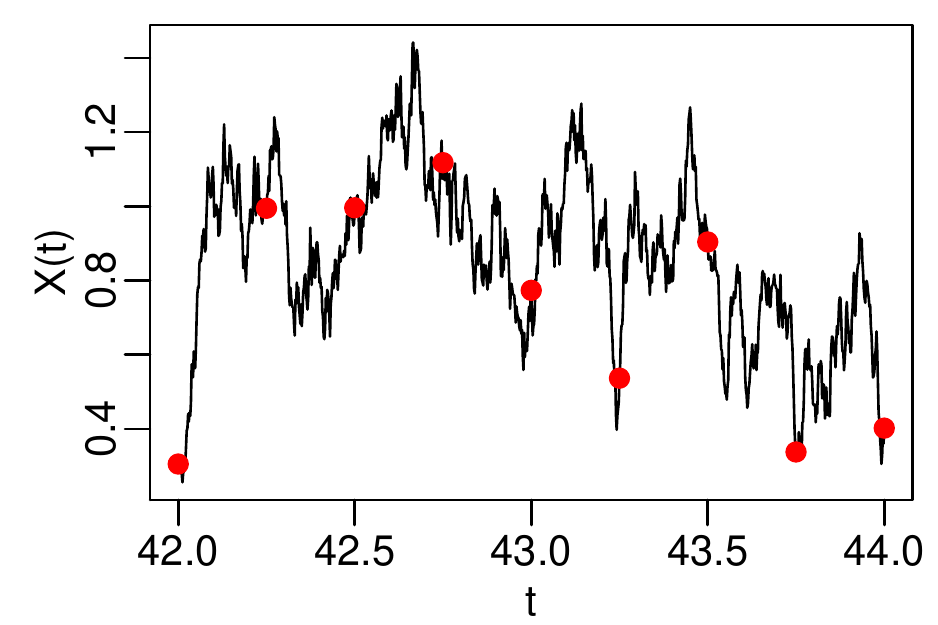}
  \caption{(color online) Snippet of the double well sample path in
    black with observations denoted as red dots.}
  \label{fig:DoubleWellnippetPath}
\end{figure}

To deal with this problem, we treat the process $X_t$ for times $t$
between consecutive observations $k \tau < t < (k+1) \tau$ as a hidden
stochastic process with a conditional path probability given by
\begin{equation}
  p(X_{0:T} | \mathbf{z}, f) \propto p(X_{0:T} | f) \prod_{k=1}^n
  \delta(z_k - X_{k \tau}),
  \label{eq:unobspath}
\end{equation}
where $\mathbf{z}$ is the collection of observations $z_k$. We will
use an iterative method based on the EM algorithm
\cite{Dempster:1977:MLI}, in which the unobserved complete paths are
replaced by an appropriate expectation using the probability
(\ref{eq:unobspath}).

\subsection{EM algorithm}

The EM algorithm cycles between two steps
\begin{enumerate}
\item In the E-step, we compute the expected negative logarithm of the
  complete data likelihood
  \begin{equation}
    \mathcal{L}(f, p) = -\mathrm{E}_p \left[ \ln L(X_{0:T} | f)
    \right],
  \end{equation}
  where $p$ denotes the posterior $p(X_{0:T} | \mathbf{z}, f_{old})$
  for the previous estimate $f_{old}$ of the drift.
\item In the M-Step, we recompute the most likely drift function by
  the minimization
  \begin{equation}
    f_{new} = \arg \min_{f} \left(\mathcal{L}(f, p) - \ln
      P_0(f) \right).
      \label{eq:mstepmin}
  \end{equation}
\end{enumerate}
On can show \cite{Dempster:1977:MLI} that the EM algorithm converges
to a local maximum of the log-posterior. To compute the expectation in
the E-step, we use (\ref{eq:likelihood}) and take the limit $\Delta t
\to 0$ at the end, when expectations have been computed. As $f(x)$ is
a time-independent function, this yields
\begin{eqnarray}
  &-& \mathrm{E}_p \left[ \ln L(X_{0:T} | f) \right] \nonumber \\
  &=& \lim_{\Delta t\to 0} \frac{1}{2} \sum_t \mathrm{E}_p \left[
    ||f(X_t)||^2 \right] \Delta t \nonumber \\
  &-& 2 \mathrm{E}_p \left[ \left( f(X_t), X_{t + \Delta t} - X_t
    \right) \right] \nonumber \\
  &=& \frac{1}{2} \int_0^T \mathrm{E}_p \left[ ||f(X_t)||^2 \right] -
  2 \mathrm{E}_p \left[ \left( f(X_t), g_t(X_t) \right) \right] dt
  \nonumber \\
  &=& \frac{1}{2} \int ||f(x)||^2 A(x) dx - \int \left( f(x), z(x)
  \right) dx.
  \label{eq:estep}
\end{eqnarray}
We have defined the corresponding drift conditioned on data
\begin{equation}
  g_t(x) = \lim_{\Delta t \to 0} \frac{1}{\Delta t} \mathrm{E}_p [X_{t
    + \Delta t} - X_t | X_t = x],
\end{equation}
as well as the functions
\begin{equation}
  A(x) = \int_0^T q_t(x) dt
  \label{eq:Afunction}
\end{equation}
and
\begin{equation}
  b(x)\, = \int_0^T g_t(x) q_t(x) dt.
  \label{eq:Bfunction}
\end{equation}
In contrast to (\ref{eq:direct}), expectations are now over marginal
densities $q_t(x)$ of $X_t$ computed from the conditional path
measure, not over the asymptotic stationary density. Hence, we end up
again with a simple quadratic form in $f$ to be minimized. Note that
due to the smoothness of the kernel the prediction of
(\ref{eq:mstepmin}) can be easily differentiated analytically, a fact
that will be needed later.

However, there are two main problems for a practical realization of
this EM algorithm:
\begin{itemize}
\item We can not compute the expectation with respect to the
  conditional path measures exactly and need to find approximations
  applicable to \emph{arbitrary} prior drift functions $f(x)$.
\item Although real observations are sparse, the hidden path involves
  a continuum of values $X_t$. This will require (e.g.~after some fine
  discretization of time) the inversion of large matrices in
  (\ref{eq:dense}).
\end{itemize}
We can readily deal with the latter problem by resorting to the sparse
GP representation introduced in section \ref{sec:sparseGP}.

\subsubsection*{Linear drift approximation: The Ornstein-Uhlenbeck
  bridge}

In this section we will look at the first problem of computing
expectations in the E-step. For given drift $f(\cdot)$ and times $t\in
I_k$ in the interval $I_k = [k \, \tau; (k + 1) \tau]$ between two
consecutive observations, the exact marginal $p_t(x)$ of the
conditional path distribution equals the density of $X_t = x$
\emph{conditioned on} the fact that $X_{k \tau} = z_k$ and $X_{(k + 1)
  \tau} = z_{k + 1}$ . This is a so-called diffusion bridge. Using the
Markov property, this density can be expressed by the transition
densities $p_s(x_{t + s} | x_t)$ of the homogeneous Markov diffusion
process with drift $f(x)$ as
\begin{equation}
  p_t(x) \propto p_{(k + 1) \tau - t}(z_{k + 1} | x) p_{t - k \tau}(x
  | z_k) \; \mbox{for}\; t\in I_k.
\end{equation}
As functions of $t$ and $x$, the second factor fulfills a forward
Fokker-Planck equation and the first one a Kolmogorov backward
equation \cite{Gardiner:1996:HSM}. Since exact computations are not
feasible for general drift functions, we \emph{approximate} the
transition density $p_s(x | x_k)$ in each interval $I_k$ by that of a
homogeneous \emph{Ornstein-Uhlenbeck process}
\cite{Gardiner:1996:HSM}, where the drift $f(x)$ is replaced by a
local linearization. Hence, we consider the approximate process
\begin{equation}
  dX_t = [f(z_k) - \Gamma_k (X_t - z_k)] dt + D^{1/2}_k dW
\end{equation}
with $\Gamma_k = -\nabla f(z_k)$ and $D_k = D(z_k)$ for $t \in I_k$.
For this process, the transition density is a multivariate Gaussian
\begin{equation}
  q_s^{(k)}(x | z) = \mathcal{N} \left( x | \alpha_k + e^{-\Gamma_k s}
    (z - \alpha_k); S_s \right),
\end{equation}
where $\alpha_k = z_k + \Gamma_k^{-1} f(z_k)$ is the stationary mean.
The covariance $S_s = A_s B_s^{-1}$ is calculated in terms of the
matrix exponential
\begin{equation}
  \left[ A_s \atop B_s \right] = \exp \left( \left[
      \begin{array}{cc}
        \Gamma_k & D_k \\
        0 & -\Gamma_k^\top
      \end{array}
    \right] s \right) \left[ 0 \atop \mathbf{I} \right].
\end{equation}
Then we obtain the Gaussian approximation $q_t^{(k)}(x) =
\mathcal{N}(x | m(t); C(t))$ of the marginal posterior for $t\in I_k$
by multiplying the two transition densities, where
\begin{eqnarray*}
  C(t) &=& \left( e^{-\Gamma_k^\top (t_{k+1} - t)} S_{t_{k+1} -
      t}^{-1} e^{-\Gamma_k (t_{k+1} - t)} + S_{t - t_k}^{-1}
  \right)^{-1}, \\
  m(t) &=& C(t) \, e^{-\Gamma_k^\top (t_{k+1} - t)} S_{t_{k+1} -
    t}^{-1} \left( z_{k+1} - \alpha_k \right. \\
  &\qquad & + \left. e^{-\Gamma_k (t_{k+1} - t)} \alpha_k \right) +
  C(t) \, S_{t - t_k}^{-1} \\
  && \left( \alpha_k + e^{-\Gamma_k (t -
      t_k)} (z_k - \alpha_k) \right).
\end{eqnarray*}
By inspecting mean and variance we see that the distribution is in
fact equivalent to a bridge between the points $X = z_k$ and $X =
z_{k+1}$ and collapses to point masses at these points.

Finally, in this approximation we obtain for the conditional drift
\begin{eqnarray*}
  g_t(x) &=& \lim_{\Delta t \to 0} \frac{1}{\Delta t} \mathrm{E}
  \left[ X_{t + \Delta t} - X_t | X_t = x, X_\tau = z_{k+1} \right] \\
  &=& f(z_k) - \Gamma_k (x - z_k) + D_k e^{-\Gamma_k^\top (t_{k+1} -
    t)} S_{t_{k+1} -t}^{-1} \\
  &&(z_{k+1} - \alpha_k - e^{-\Gamma_k (t_{k+1} - t)} (x - \alpha_k))
\end{eqnarray*}
as shown in appendix \ref{sec:pdrift}.

\subsubsection*{Sparse M-Step approximation}

For the M-Step approximation we use the sparse GP formalism of section
\ref{sec:sparseGP}. The resulting sparse approximation to the
likelihood (\ref{eq:estep}) is given by
\begin{eqnarray}
  \mathcal{L}_s(\mathbf{f},q) &=& \frac{1}{2} \int ||
  \mathrm{E}_0[f(x) | \mathbf{f}_s] ||^2 \; A(x) \; dx \nonumber \\
  &\qquad& - \int \left (\mathrm{E}_0[f(x) |
    \mathbf{f}_s], b(x) \right ) \; dx,
\end{eqnarray}
where the conditional expectation is over the GP prior. While the
exact likelihood does not contain interactions of the form $f(x)f(x')$
for $x\neq x'$, we allow for couplings of the type $\frac{1}{2}
\mathbf{f}^\top \boldsymbol{\Lambda} \mathbf{f} - \mathbf{a}^\top
\mathbf{f}$ in the effective log-likelihood.

In order to avoid cluttered notation, it should be noted that in the
following results for a component $f^j$, the quantities
$\boldsymbol{\Lambda}_s, \mathbf{f}_s, \mathbf{k}_s,
\mathbf{K}_s^{-1}, z(x), \mathbf{D}(x)$ similar to (\ref{eq:dense})
depend on the component $j$, but not $A(x)$.

We easily get
\begin{equation}
  \mathrm{E}_0[f(x) | \mathbf{f}_s] = \mathbf{k}_s^\top(x)
  \mathbf{K}_s^{-1} \mathbf{f}_s.
\end{equation}
Hence
\begin{equation}
  \mathcal{L}_s(\mathbf{f},q) = \frac{1}{2} \mathbf{f}_s^\top
  \boldsymbol{\Lambda}_s \mathbf{f}^s - \mathbf{f}_s^\top \mathbf{y}_s
\end{equation}
with
\begin{equation}
  \boldsymbol{\Lambda}_s = \mathbf{K}_s^{-1} \left \{ \int
    \mathbf{k}_s(x)  \mathbf{D}(x)^{-1} \; A(x) \;
    \mathbf{k}_s^\top(x) dx \right\} \mathbf{K}_s^{-1}
  \label{eq:sparse1}
\end{equation}
and
\begin{equation}
  \mathbf{y}_s \, =  \mathbf{K}_s^{-1} \int
  \mathbf{D}(x)^ {-1}\mathbf{k}_s(x) \; b(x)\; dx.
  \label{eq:sparse2}
\end{equation}
With these results, the approximate \emph{MAP} estimate is
\begin{equation}
  \bar{f}_s(x) = \mathbf{k}^\top_s(x)(\mathbf{I} +
  \boldsymbol{\Lambda}_s \mathbf{K}_s)^{-1} \mathbf{y}_s.
\end{equation}
The integrals over $x$ in (\ref{eq:sparse1}) and (\ref{eq:sparse2})
can be computed analytically for many kernels of interest such as
polynomial and RBF ones. However, we found it more efficient to treat
the time integration in (\ref{eq:Afunction}) and (\ref{eq:Bfunction})
as well as the $x$-integrals by sampling, where time points $t$ are
drawn uniformly at random and $x$ points from the multivariate
Gaussian $q_t(x)$. A related expression for the variance,
\begin{equation}
  \bar{D}_s(x) = K(x, x) - \mathbf{k}^\top_s(x) (\mathbf{I} +
  \boldsymbol{\Lambda} \mathbf{K}^s)^{-1} \boldsymbol{\Lambda}_s
  \mathbf{k}_s(x),
\end{equation}
can only be viewed as a crude estimate, because it does not include
the impact of the GP fluctuations on the path probabilities.

Finally, a possible approximate evidence for our model is given by the
product of the local Ornstein-Uhlenbeck transition probabilities:
\begin{equation}
  p(\mathbf{z}) \approx p_{ou}(\mathbf{z}|\hat{\mathbf{f}}) = p(x_1)
  \prod_{j=1}^{n-1} q^{(k)}_{\tau} (z_{k+1} | z_{k}).
  \label{eq:approxfree}
\end{equation}
The expression is a product of Gaussian transition densities and
therefore of analytical form. Note that in addition to the
Ornstein-Uhlenbeck linearization, this approximation also neglects the
uncertainty of $\mathbf{f}$, since the GP in the M step only uses the
expectation.

Nevertheless, in our experiments we found that the use of the
approximate evidence is a reasonable choice for the optimization of
the diffusion $D(x)$, see subsection \ref{subsec:diffest}. However,
the optimization of the kernel hyperparameters is more problematic,
since the approximate evidence depends on the drift estimate
$\hat{\mathbf{f}}$, which itself depends on the choice of the
hyperparameters through the application of the GP. Since we assume
that prior knowledge of a suitable kernel hyperparameters is often
available, we did not pursue this problem further.

\subsection{Experiments}
\label{sec:experiments}

We created the synthetic data sets in this section by first using the
Euler method from the corresponding SDE with grid size
$\Delta_{\mathrm{dense}}=0.002$. Then for a data set of $N$
observations separated by $\Delta t \gg \Delta_{\mathrm{dense}}$, we
keep every $k = (\Delta t / \Delta_{\mathrm{dense}})$th path sample
value as observation, until the desired observation number $N$ is
reached.

The EM algorithm is initialized with the sparse direct GP estimator,
which works well in practice as a reasonable first approximation to
the true system dynamics. Although the monotonicity property of the EM
algorithm is no longer satisfied due to the approximation in the
E-step, convergence will be assumed, once $\mathcal{L}$ stabilizes up
to some minor fluctuations. In our experiments convergence was
typically attained after a few ($<10$) iterations.

\subsubsection*{Performance Comparison}

First, we compare the estimation accuracy of the direct GP and the EM
algorithm on the double well model with constant known diffusion,
\begin{equation}
  dX = 4(X - X^3) dt + dW_t,
\end{equation}
for different time discretization $\Delta t$. For each time step, we
generated $20$ data sets, each of size $n = 4000$, and computed the
MSE on a test set of size $n = 2000$ for each data set for both
algorithms with RBF kernel. As benchmark reference, we include the
estimation results of a Monte Carlo sampler (see appendix
\ref{sec:mcmc}). The latter one is represented only for one data set
at small and medium time intervals, respectively, due to its long
computation time. In order to improve comparability, we fixed the
length scale of the RBF kernel to $l=0.62$ for all data sets.

\begin{figure}
  \centering
  \includegraphics[width=0.4\textwidth]{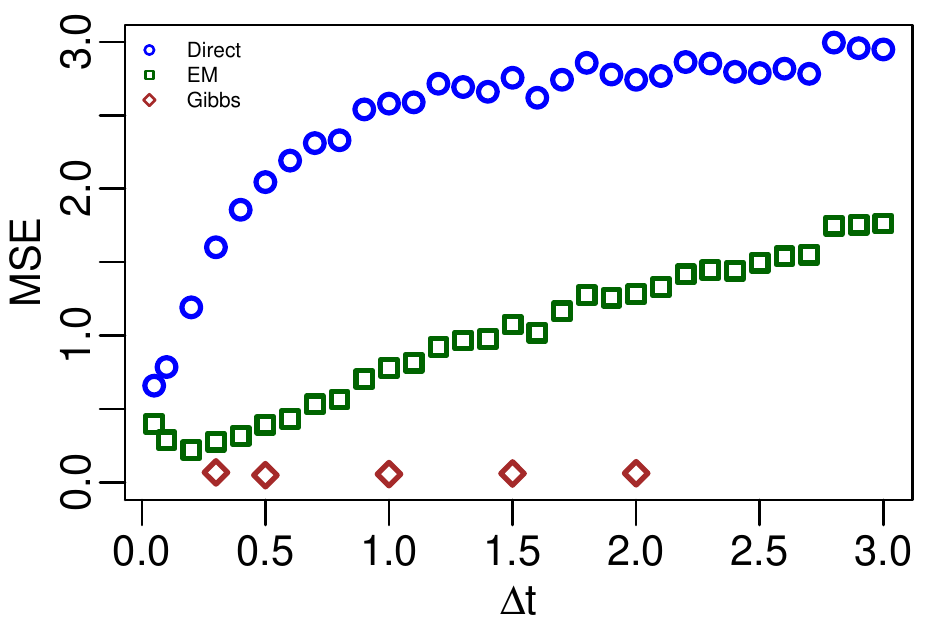}
  \caption{(color online) Comparison of the MSE for different methods
    over different time intervals.}
  \label{fig:msecomparison}
\end{figure}

The results are given in figure \ref{fig:msecomparison}. The MSE of
the direct GP grows quite rapidly for smaller intervals until it
reaches an upper bound roughly equivalent with randomly guessing the
drift function. On the other hand, the MSE for the EM algorithm
increases at a much slower rate, giving good results even for data
sets with bigger time distances. The estimation results for the Gibbs
sampler are independent of the discretization rate, but take
considerable time to compute: while the EM algorithm runs for a couple
of minutes, the sampler takes up to two days.

\subsubsection*{Double well model with known state dependent
  diffusion}

\begin{figure}
  \centering
  \includegraphics[width=0.4\textwidth]{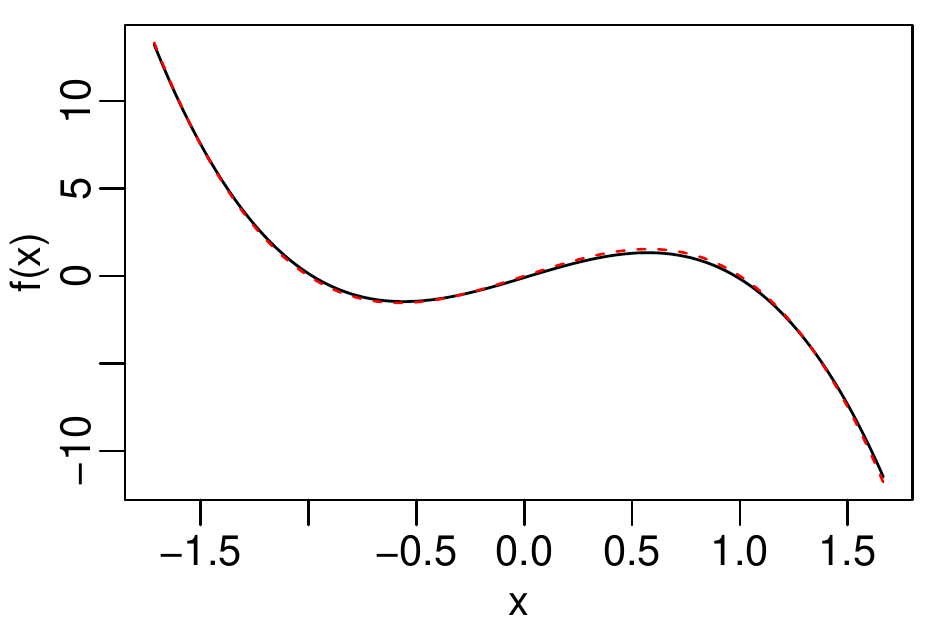}
  \caption{(color online) GP estimation after one iteration of the EM
    algorithm. Again, the solid black and red dashed lines denote
    estimator and true drift function, respectively.}
  \label{fig:KnownDWDiffusionEM}
\end{figure}

As our next example we examine the double well model with state
dependent diffusion and larger time discretization. Here we assume
that the diffusion function $D(x)$ is known. Specifically, we sample
$n = 4000$ observation at $\Delta t = 0.5$ and run the EM algorithm
with a polynomial kernel of order $p=4$. The direct GP and the EM
result are given in figure \ref{fig:WrongDrift} and
\ref{fig:KnownDWDiffusionEM}, respectively. One can clearly see, that
an application of the EM algorithm leads to a significantly better
estimator of the drift function, compared to the direct GP method.

\subsubsection*{Two dimensional synthetic model}

\begin{figure}
  \centering
  \includegraphics[width=0.4\textwidth]{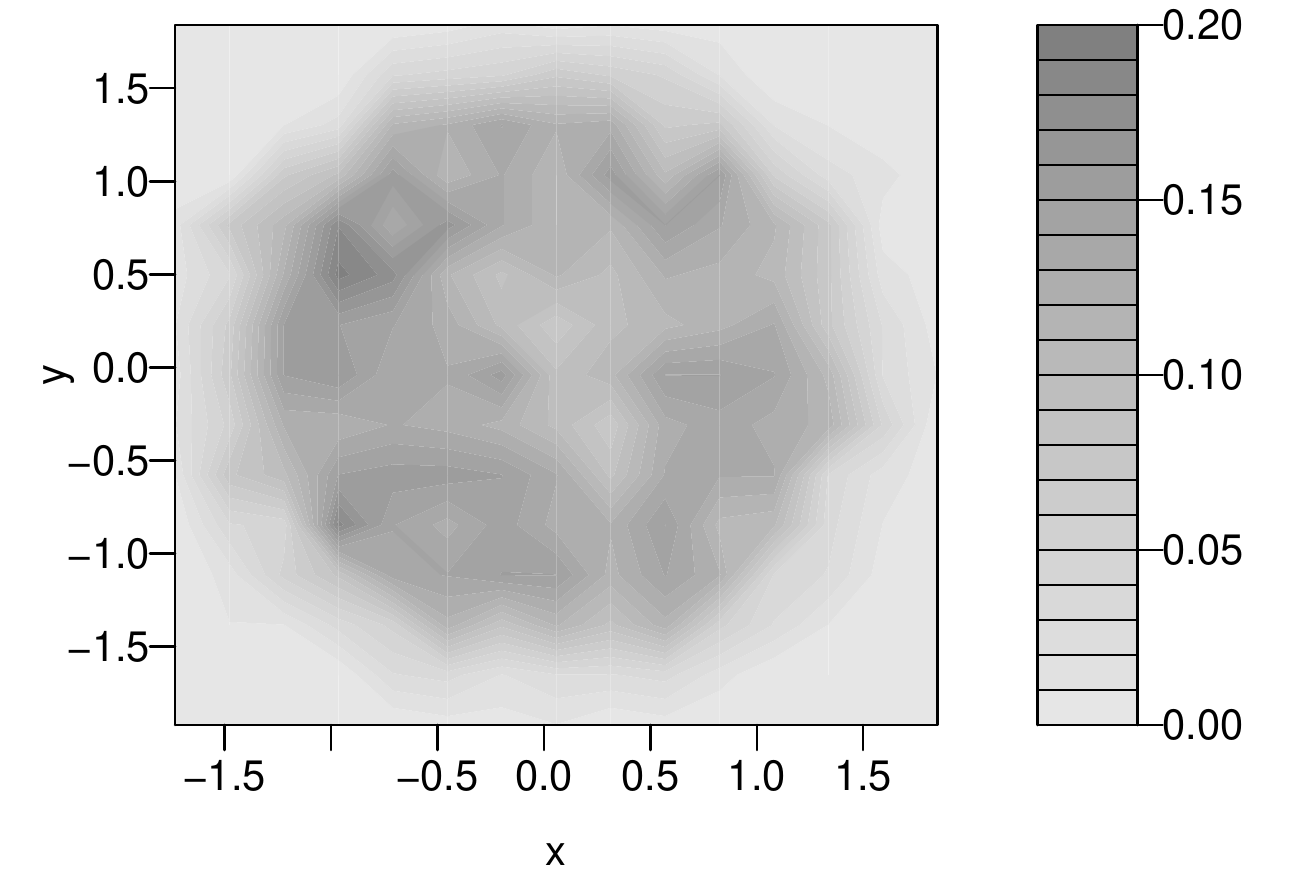}
  \caption{Empirical density of the two dimensional synthetic model
    data.}
  \label{fig:DW2DHist}
\end{figure}

\begin{figure}
  \centering
  \includegraphics[width=0.4\textwidth]{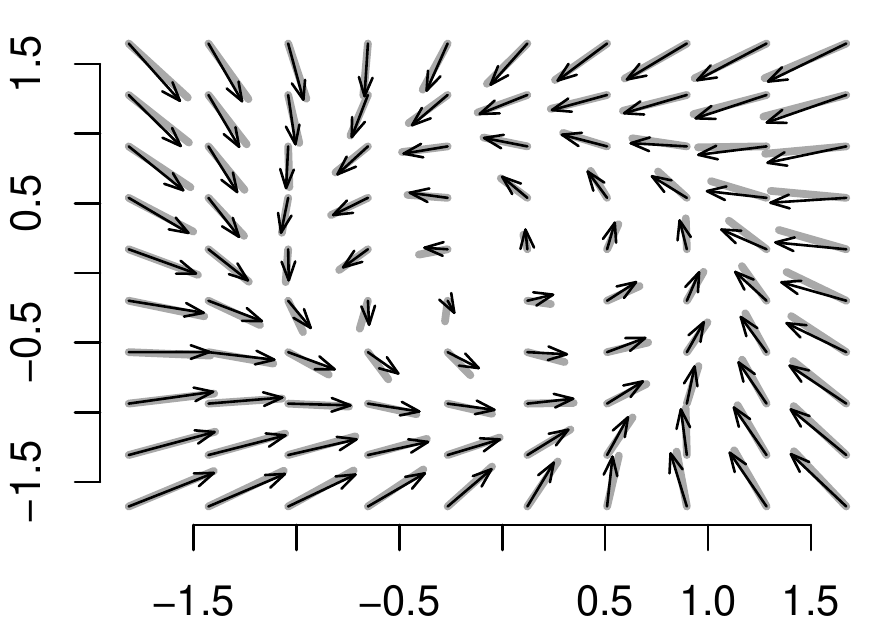}
  \caption{Vector fields of the true drift depicted as grey lines and
    the estimated drift as black arrows for the two dimensional
    synthetic data.}
  \label{fig:DW2DEst}
\end{figure}

We now turn to a two dimensional process with the following dynamics:
\begin{eqnarray}
  dX &=& (X (1 - X^2 - Y^2) - Y)dt + dW_t^{(1)}, \\
  dY &=& (Y (1 - X^2 - Y^2) + Ý)dt + dW_t^{(2)},
\end{eqnarray}
where the component indices are denoted by superscripts. For this
model we generated $n = 10000$ observations with step size $\Delta t =
0.2$ shown in figure \ref{fig:DW2DHist}. The estimation in figure
\ref{fig:DW2DEst} uses a polynomial kernel of order $p = 4$ and shows
a good fit to the true drift especially in the regions where the
observations are concentrated. Note that this is a non-equilibrium
model, where the drift cannot be expressed as the gradient of a
potential. Hence, the density based method of \cite{Batz:2016:VED}
cannot be applied here.

\subsubsection*{Lorenz'63 model}

We next analyze a stochastic version of the three dimensional
Lorenz'63 model. It consists of the following system of nonlinear
coupled stochastic differential equations:
\begin{eqnarray}
  dX &=& \sigma (Y - X)dt + dW_t^{(1)}, \\
  dY &=& (\rho X - X - XZ)dt + dW_t^{(2)}, \\
  dZ &=& (XY - \beta Z)dt + dW_t^{(3)}.
\end{eqnarray}
Lorenz'63 is a chaotic system which was developed as a simplified
model of thermal convection in the atmosphere \cite{Lorenz:1963:DNF}.
The parameters $\boldsymbol{\theta} = (\sigma, \rho, \beta)$ are set
to the commonly used $\boldsymbol{\theta} = (10, 28, 8 / 3)$ known to
induce chaotic behavior in the system. In order to analyze the model
we simulate $n = 3000$ data points with time discretization $\Delta t
= 0.2$. In the inference, we used a polynomial kernel of order $p = 2$
and assume that the constant diffusion is known.

\begin{figure}
  \centering
  \includegraphics[width=0.4\textwidth]{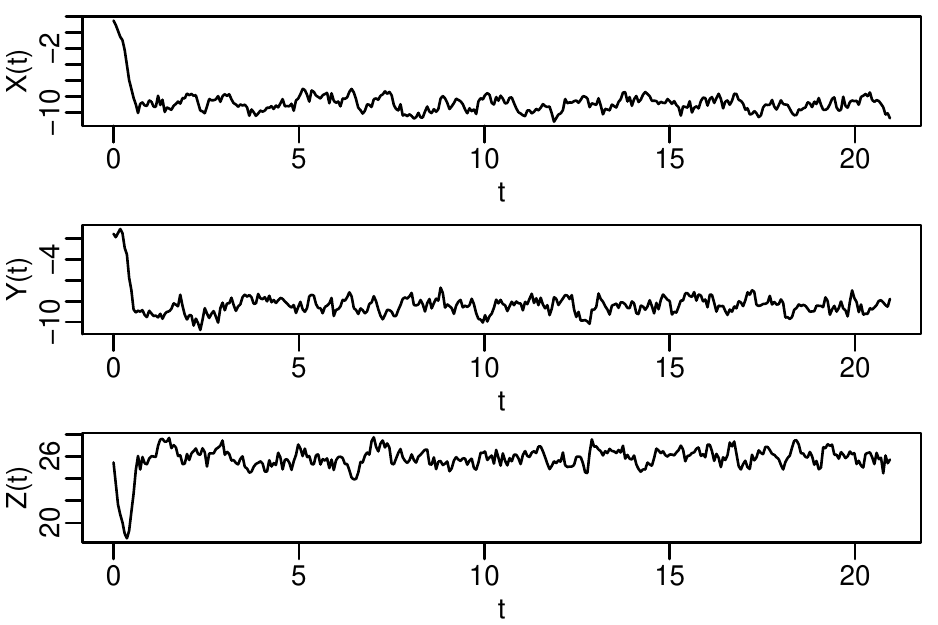}
  \caption{Simulated sample path of the Lorenz'63 model learned by the
    direct GP algorithm.}
  \label{fig:Lorenz63_1}
\end{figure}

\begin{figure}
  \centering
  \includegraphics[width=0.4\textwidth]{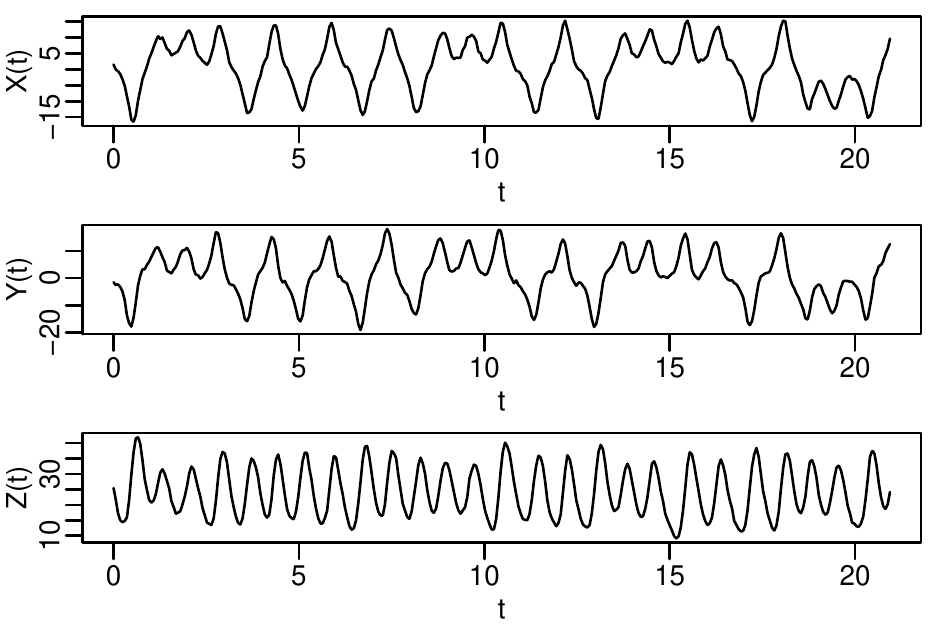}
  \caption{Simulated sample path of the Lorenz'63 model learned by the
    EM algorithm.}
  \label{fig:Lorenz63_2}
\end{figure}

\begin{figure}
  \centering
  \includegraphics[width=0.4\textwidth]{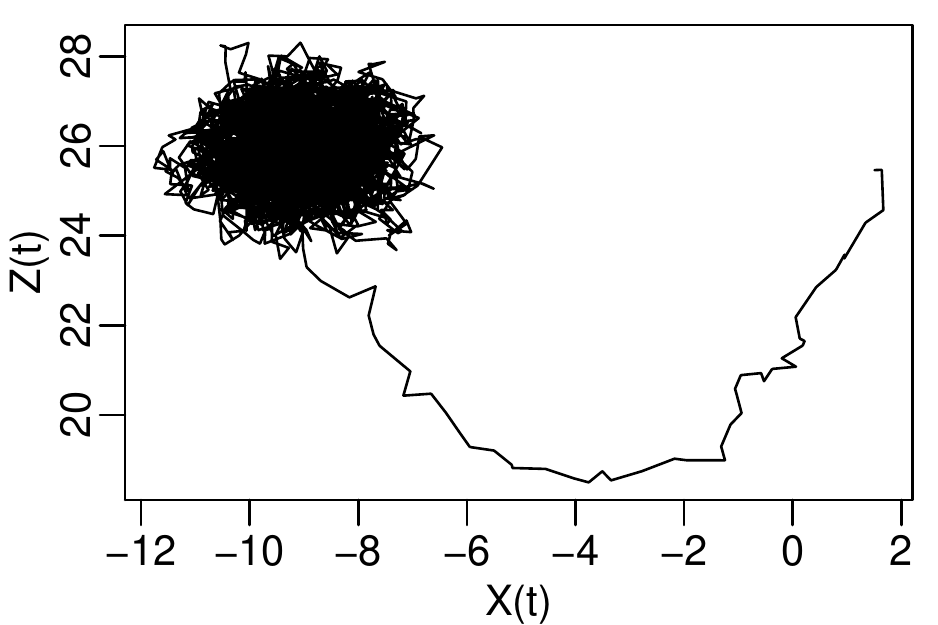}
  \caption{Simulated path in the X-Z plane from the Lorenz'63 model
    learned by the direct GP algorithm.}
  \label{fig:Lorenz63_3}
\end{figure}

\begin{figure}
  \centering
  \includegraphics[width=0.4\textwidth]{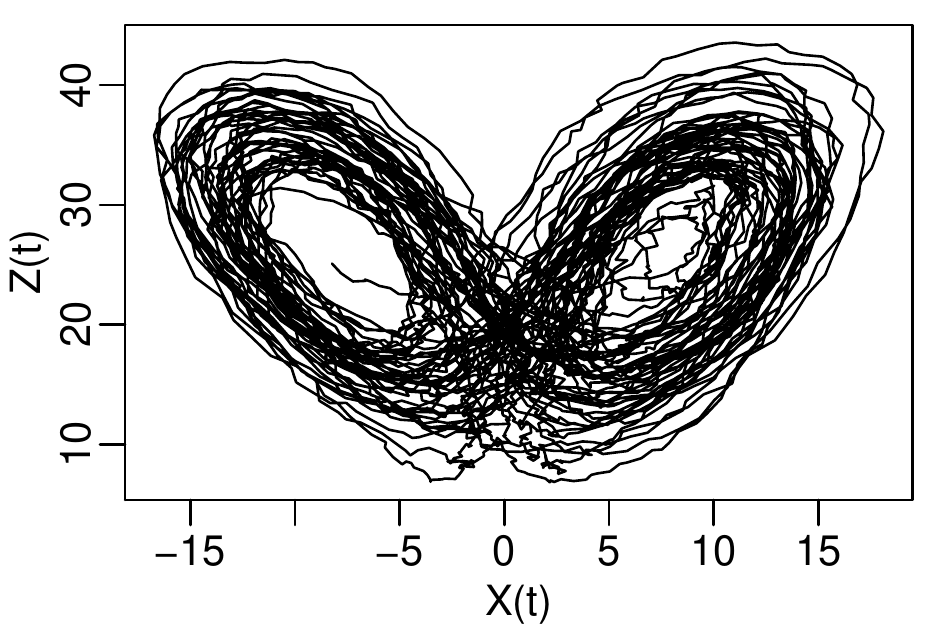}
  \caption{Simulated path in the X-Z plane from the Lorenz'63 model
    learned by the EM algorithm.}
  \label{fig:Lorenz63_4}
\end{figure}

In order to visualize the quality of the estimation results, we
computed the direct GP and the EM algorithm and simulated paths using
the corresponding mean estimator as drift function. Here, the
application of the EM leads to a vastly superior estimation result
compared to the direct method. As shown in figure
\ref{fig:Lorenz63_3}, the direct GP estimator path collapses to a
small region of the function space, whereas the EM trajectory of
figure \ref{fig:Lorenz63_4} nicely captures the true dynamics of the
Lorenz'63 model, faithfully recreating the famous butterfly pattern in
the X-Z plane.

\subsubsection*{Cart and pole model}

\begin{figure}
  \centering
  \includegraphics[width=0.4\textwidth]{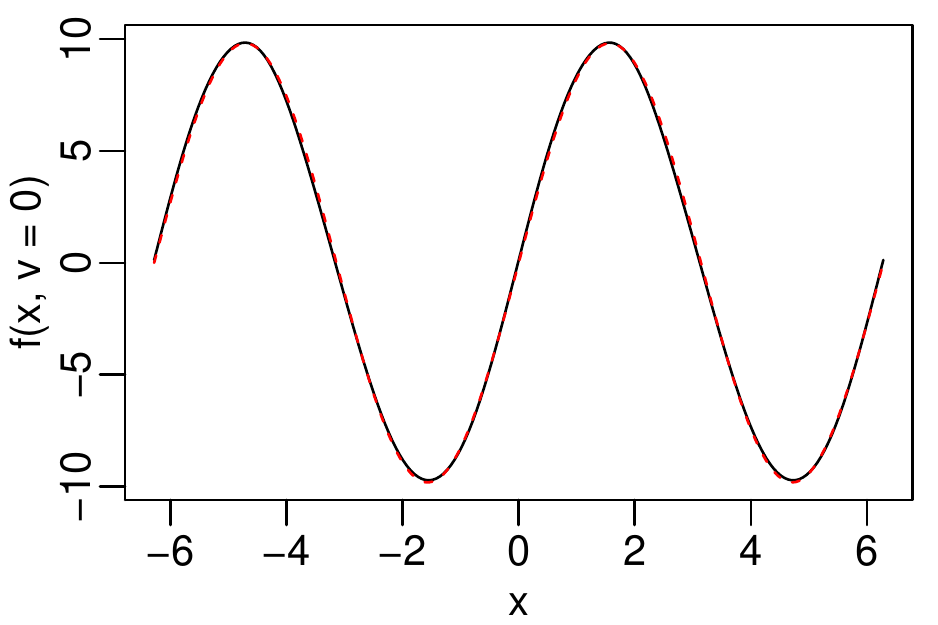}
  \caption{(color online) One dimensional drift estimation plots of
    the second ($dV$) SDE of the cart and pole model. The figure shows
    the estimation of the pendulum position $X$ for a fixed velocity
    $V = 0$. The solid black line is the drift estimation and the red
    dashed line the true function.}
  \label{fig:CartPoleContour1}
\end{figure}

\begin{figure}
  \centering
  \includegraphics[width=0.4\textwidth]{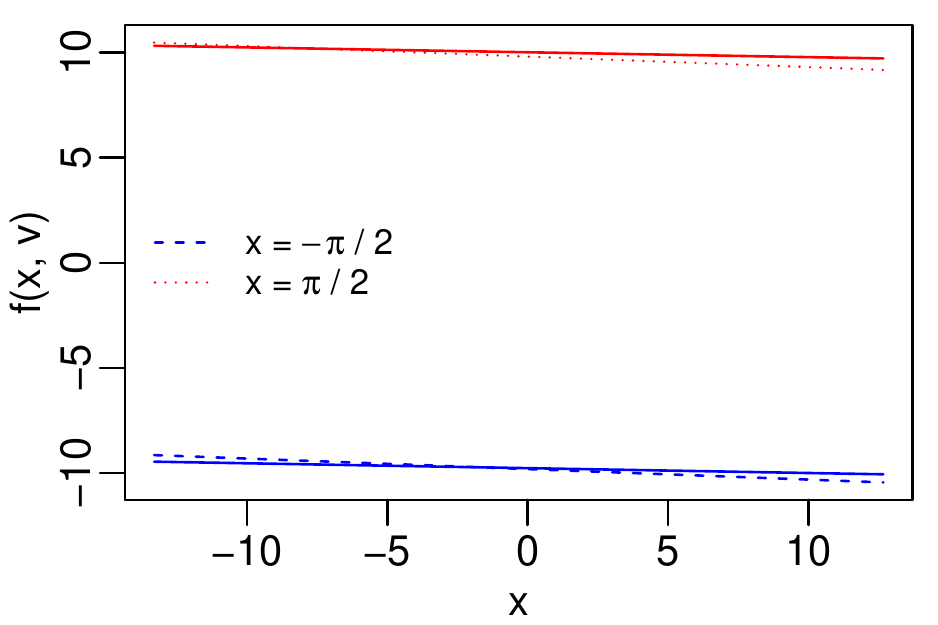}
  \caption{(color online) One dimensional drift estimation plots of
    the second ($dV$) SDE of the cart and pole model: Estimation of
    $V$ for a fixed pendulum position in the horizontal positions with
    the top pointing to the left $X = -\pi / 2$ and to the right side
    $X = \pi / 2$. Full lines denote the drift estimation and dashed
    and dotted lines the true values.}
  \label{fig:CartPoleContour2}
\end{figure}

Next, we consider an example from the class of mechanical systems. Our
model describes the dynamics of a pole attached to a cart moving
randomly along an one-dimensional axis. Formally, we get a system of
two dimensional differential equations with $x$ denoting the angle of
the pendulum, and $v$ the angular velocity. We define the upright
position of the pendulum as $X = 0$. This particular \emph{cart and
  pole} model is frequently studied in the context of learning control
policies \cite{Deisenroth:2009:GPD}, where the goal is to move the
cart in such a way as to stabilize the pendulum in the upright
position. The complete system looks as follows:
\begin{eqnarray}
  dX &=& Vdt, \\
  dV &=& \frac{-\gamma V + mgl \sin(X)}{ml^ 2}dt + d^{1/2} dW_t,
\end{eqnarray}
where $\gamma = 0.05$ is the friction coefficient, $l = 1 \mathrm{m}$
and $ m = 1 \mathrm{kg}$ are the length and mass of the pendulum,
respectively, and $g=9.81 \mathrm{m} \, \mathrm{s}^{-2}$ denotes the
gravitational constant. For our experiment, we generated $N=4000$ data
points $(x, v)$ on a grid with $\Delta t = 0.3$ and known diffusion
constant $d = 1$. Here, the full diffusion matrix
\begin{equation}
  D = \left(
    \begin{array}{cc}
      0 & 0 \\
      0 & 1
    \end{array}
  \right),
\end{equation}
for both $X$ and $V$ is rank deficient due to its noiseless first
equation. However, we note that our EM algorithm is also applicable to
models with deterministic components, since the E-Step in the EM
algorithm remains well defined. In the kernel function we incorporate
our prior knowledge that the pendulum angle is periodic and the
velocity acts as a linear friction term inside the system.
Specifically we define the following multiplicative kernel for the
$dV$ equation:
\begin{equation}
  K\left((x, v), (x', v')\right) = K_\mathrm{Per}(x, x')
  K_\mathrm{Poly}(v, v'),
\end{equation}
where $K_\mathrm{Per}$ denotes the periodic kernel over the state $x$
with hyperparameters $l=1.21$ and $K_\mathrm{Poly}$ the polynomial
kernel of order $p=1$ over the velocity $V$. The multiplicative kernel
structure allows for interactions between its components. Since in
this model the components are independent, we could also use an
additive kernel, which neglects interactions terms, but we have chosen
the more generally applicable variant here. For the $dX$ equation, we
use a polynomial kernel of order $p=1$, which captures the linear
relationship between $X$ and $V$. If we adapt our choice of the kernel
to the specific form of the system, we get an accurate estimate even
for data points separated by a wider time spacing (see figures
\ref{fig:CartPoleContour1} and \ref{fig:CartPoleContour2}).

\subsection{External forces}

We can expect a reasonably good estimation of $f(x)$ only in regions
of $x$ where we have enough observations. This is of clear importance,
when the system is multi-stable and the noise is too small to allow
for a sufficient exploration of space. An alternative method for
exploration would be to add a known external \emph{deterministic
  control force} $u(t)$ to the dynamics which is designed to drive the
system from one locally stable region to another one. Hence, we assume
an SDE
\begin{equation}
  dX_t = \left(f(X_t) + u(t)\right) dt + D^{1/2} dW_t.
  \label{eq:SDE_control}
\end{equation}
This situation is easily incorporated into our formalism. In all
likelihood terms, we replace $f(X_t)$ by $f(X_t) + u(t)$, but keeping
the zero mean GP prior over functions. The changes for the
corresponding transition probabilities of the approximating time
dependent Ornstein-Uhlenbeck bridge are given in appendix
\ref{sec:xou}.

\begin{figure}
  \centering
  \includegraphics[width=0.4\textwidth]{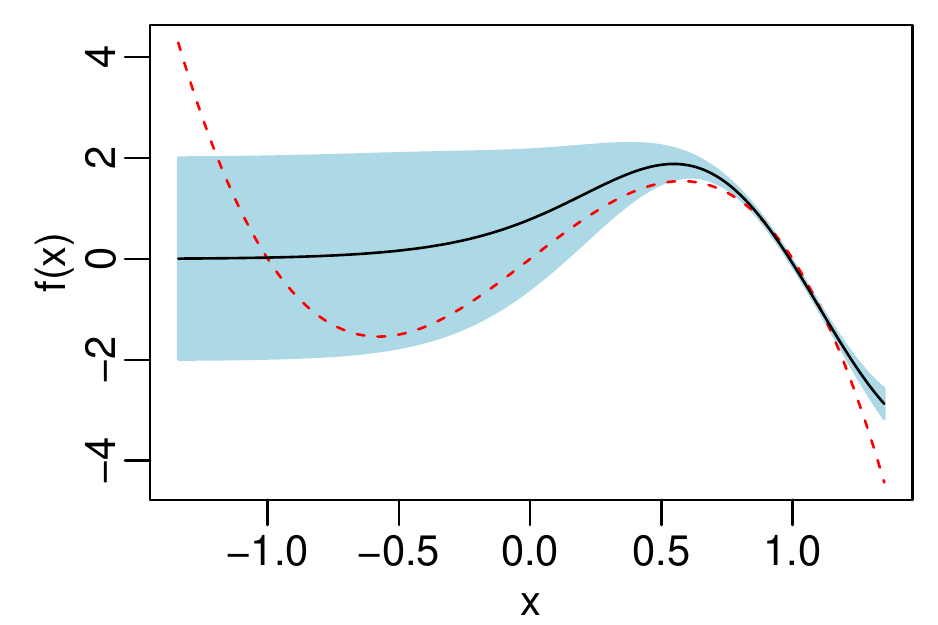}
  \caption{(color online) EM algorithm predictions for the
    uncontrolled double well path with the solid black line denoting
    the estimation and the dashed red line the true drift. Here, the
    estimation of the well around $X=-1$ basically equals the GP
    prior, since there are no observations on this region. The shaded
    area can be interpreted as the 95\%-confidence bound.}
    \label{fig:DWControlEst_1}
\end{figure}

\begin{figure}
  \centering
  \includegraphics[width=0.4\textwidth]{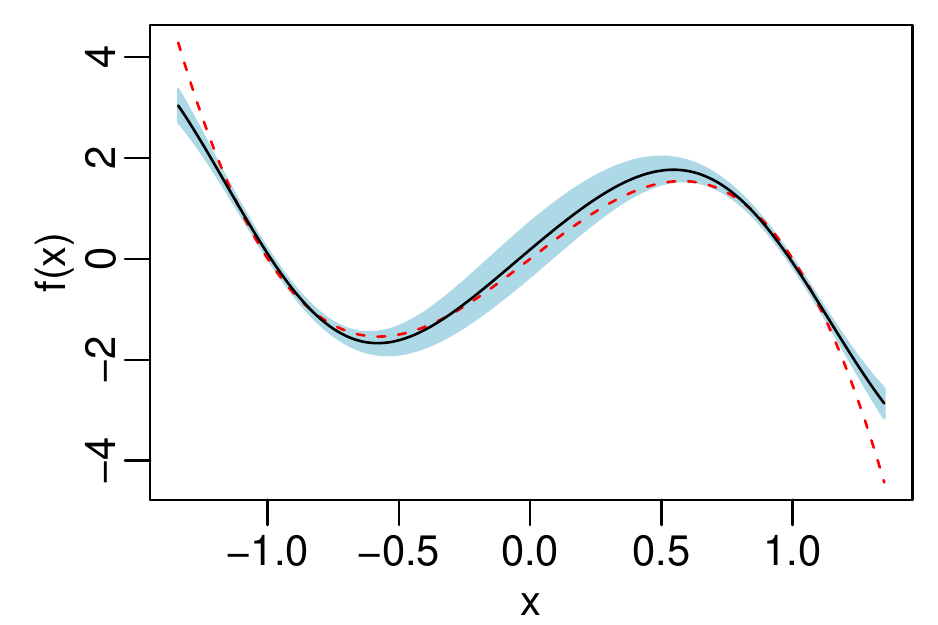}
  \caption{(color online) EM algorithm predictions for the controlled
    double well path. The solid black line is the estimated drift and
    the dashed red line the true function.}
  \label{fig:DWControlEst_2}
\end{figure}

We demonstrate the concept by applying it to the double well model. We
get
\begin{equation}
  dX = (4(X - X^3) + u(t))dt + \sigma dW_t.
\end{equation}
As external force we choose a periodic control function of the form
$u(t) = a\sin(\omega t)$ with parameters $a=1$ and $\omega = 3$. We
generated a data set of $n = 2000$ observations on a regular grid with
distance $\Delta t = 0.2$ from the model with known diffusion $D^{1/2}
= 0.5$. The addition of $u(t)$ leads to observations from both of the
wells, whereas in the uncontrolled case only one part of the
underlying state space is explored. Hence, the drift estimation in the
latter case leads to an accurate result solely around the well at
$X=1$, as opposed to the controlled case, where both modes are
truthfully recovered (figures \ref{fig:DWControlEst_1} and
\ref{fig:DWControlEst_2}). In both cases, we used a RBF kernel with
$\tau=1$. The length scales was set to $l = 0.74$ in the controlled
and $l = 0.53$ in the uncontrolled case.

\subsection{Diffusion Estimation}
\label{subsec:diffest}

As in the dense data scenario, we look at constant and state dependent
diffusions in turn. If $D$ does not depend on the state, we can
proceed in analogy to the dense data case and maximize the approximate
evidence (\ref{eq:approxfree}) with respect to the diffusion values.

For the state dependent case $D(x)$ we assume a parametric function
$D(x;\theta)$, which is specified by its parameter vector $\theta$.
Here, we again maximize the likelihood with respect to the
corresponding $\theta$.

For an illustration, we don't show the constant diffusion case and
instead restrict ourself to the more interesting case of a state
dependent $D(x)$. We sampled $n = 8000$ observations at $\Delta t =
0.3$ from the following process:
\begin{equation}
  dX = 0.4(4 - X) dt + \max(2-(X - 4)^2, 0.25)dW_t.
  \label{eq:NPDiffusionSDE}
\end{equation}

\begin{figure}
  \centering
  \includegraphics[width=0.4\textwidth]{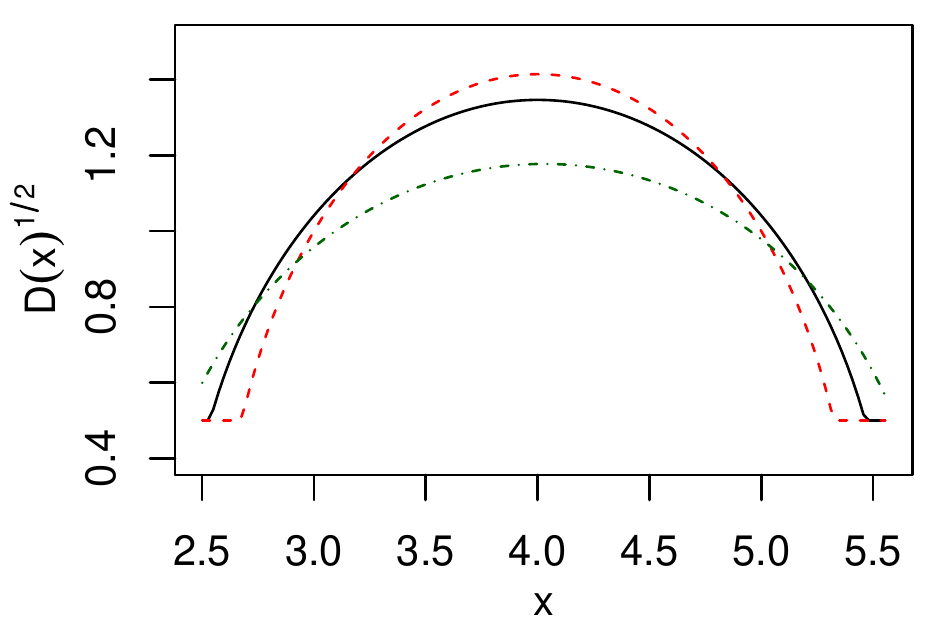}
  \caption{(color online) Comparison of the diffusion estimation for
    data generated from (\ref{eq:NPDiffusionSDE}). The dashed red line
    is the true square root $D(x)^{1/2}$ of the diffusion and the
    solid black line the parametric estimation based on the EM
    algorithm. For comparison, we include the estimate based on the
    direct GP denoted by the green dashed-dotted line.}
  \label{fig:ComparisonOUDiffusion}
\end{figure}

The diffusion function was modelled as $D(x, \mathbf{\theta}) =
\theta_1 x^2 + \theta_2 x + \theta_3$. As kernel function for the
drift, we used a polynomial kernel of order $p = 1$. Optimizing the
evidence with respect to $\bf{\theta}$ leads to the results shown in
figure \ref{fig:ComparisonOUDiffusion}. One can see that the
estimation gives a reasonably good fit to the true diffusion function
even with the bigger time discretization. We note however, that the
diffusion estimate is of a lower quality than the drift estimate,
since in this case the evidence is less accurate.

\section{Discussion}
\label{sec:discussion}

It would be interesting to replace the ad hoc local linear
approximation of the posterior drift by a more flexible time dependent
Gaussian model. This could be optimized in a variational EM
approximation by minimizing a free energy in the E-step, which
contains the Kullback-Leibler divergence between the linear and true
processes \cite{Archambeau:2007:VID, Vrettas:2015:VMF}. Such a method
could be extended to noisy observations and the case, where some
components of the state vector are not observed. Also, this method
could be turned into a variational Bayesian approximation, where one
optimizes posteriors over both drifts and over state paths. The path
probabilities are then influenced by the uncertainties in the drift
estimation, which would lead to more realistic predictions of error
bars.

Finally, nonparametric diffusion estimation deserves further
attention. Incorporating a fully nonparametric model of the diffusion
function $D(x)$ in our scheme would be infeasible in practice, since
this would involve the joint estimation of $n$ diffusion matrices. In
our experiments, we tried a (quasi-)nonparametric approach, where we
represented the diffusion function by its value at a few supporting
points and took these as inputs for a GP regression, which we then
used as function approximation. However, our experiments have shown
that in order to achieve a reasonable estimation quality we need
supporting points on a relatively dense grid. The corresponding
optimization over the vector of grid points turned out to be too
inefficient, which makes the approach impractical. Furthermore, the
evidence over which we optimize is often too inaccurate to lead to a
reasonable quality.

If performance time is not at all critical, one can resort to a Markov
Chain Monte Carlo (MCMC) algorithm, which generates exact samples from
the corresponding drift and diffusion functions. In contrast to the EM
algorithm, the sampler evaluates the diffusion function on a dense
grid and also does not use the assumption of constant diffusion
between adjacent observations, thereby overcoming the significant
estimation errors for larger time distances. We plan to report on this
in a future publication.

\subsubsection*{Acknowledgments}

This work was supported by the European Community's Seventh Framework
Programme (FP7, 2007-2013) under the grant agreement 270327
(CompLACS).

\appendix

\begin{widetext}

\section{Conditional drift}
\label{sec:pdrift}

Here, we give the derivation of the conditional drift term $g_t(x)$,
which occurs in the E-step of the EM algorithm.

\begin{eqnarray*}
  g_t(x) &=& \lim_{\Delta t\to 0} \frac{1}{\Delta t} \mathrm{E} \left[
    X_{t + \Delta t} - X_t | X_t = x, X_\tau = y \right] \\
  &=& \lim_{\Delta t \to 0} \frac{1}{\Delta t} \frac{\int (x' - x) \;
    p_{\tau - t - \Delta t}(y | x') p_{\Delta t} (x' | x) \; dx'}{\int
    p_{\tau - t - \Delta t}(y | x') p_{\Delta t} (x' | x) \; dx'} \\
  &=& \lim_{\Delta t \to 0} \frac{1}{\Delta t} \frac{f(x) \Delta t +
    \mathrm{E}_u\left[ p_{\tau - t - \Delta t}(y | x + f(x) \Delta t +
      u) u \right]}{\mathrm{E}_u\left[ p_{\tau - t - \Delta t}(y | x +
      f(x) \Delta t + u) \right]} \\
  &=& f(x) + D \lim_{\Delta t \to 0} \frac{\nabla_x \mathrm{E}_u\left[
      p_{\tau - t - \Delta t}(y | x + f(x) \Delta t + u)
    \right]}{\mathrm{E}_u\left[ p_{\tau - t - \Delta t}(y | x + f(x)
      \Delta t + u) \right]} \\
  &=& f(x) + D \lim_{\Delta t \to 0} \nabla_x \ln \left \{
    \mathrm{E}_u \left[ p_{\tau - t - \Delta t}(y | x + f(x) \Delta t
      + u) \right] \right \} \\
  &=& f(x) + D \nabla_x \ln \left \{ p_{\tau - t}(y | x) \right \}.
\end{eqnarray*}

The second line follows from the definition of the conditional
density, the 3rd line from the fact that $p_{\Delta t} (x' | x) =
{\cal{N}}( x + f(x)\Delta t ; D \Delta t)$ and $u \sim {\cal{N}}(0 ;
\sigma^2 \Delta t)$. The fourth line is based on the fact that for
zero mean Gaussian random vectors with covariance $S$, we have
$\mathrm{E}[u g(u)] = S \mathrm{E}[\nabla_u g(u)]$. Finally, the last
line is obtained by noting that the covariance of $u$ vanishes for
$\Delta t \to 0$.

\section{Ornstein-Uhlenbeck bridge with external forces}
\label{sec:xou}

If there is an additional time-dependent and known drift term $u(t)$,
e.g.~a control force, in the Ornstein-Uhlenbeck model, i.e.
\begin{displaymath}
  dX_t = [f(y_k) - \Gamma_k (X_t - y_k) + u(t)] dt + D_k^{1/2} dW,
\end{displaymath}
with $\quad \Gamma_k = -\nabla f(y_k)$ and $D_k = D(y_k)$, the mean of
the marginal posterior is changed to
\begin{align*}
  m(t) &= C(t) e^{-\Gamma_k^\top (\tau - u)} S_{\tau-u}^{-1} \left(
    x_{k+1} - \alpha_k + e^{-\Gamma_k (\tau - u)} \alpha_k -
    \int_u^\tau e^{-\Gamma_k (\tau - v)} u(t-u+v) dv \right) \\
      &+ C(t) S_u^{-1} \left( \alpha_k + e^{-\Gamma_k u} (x_k -
        \alpha_k) + \int_0^u e^{-\Gamma_k (u - v)} u(t-u+v)) dv
      \right),
\end{align*}
but the covariance matrix stays the same. For the posterior drift, we
get in this case
\begin{equation*}
  g_t(x) \approx f(x_k) - \Gamma_k (x - x_k) + D_k e^{-\Gamma_k^\top
    (\tau - u)} S_{\tau - u}^{-1} \left( x_{k+1} - \alpha_k
    - e^{-\Gamma_k (\tau - u)} (x - \alpha_k) -
      \int_u^\tau e^{-\Gamma_k (\tau - v)} u(t-u+v)) dv \right).
\end{equation*}
For $u(t) = a \sin(\omega t)$:
\begin{align*}
  m(t) &= C(t) e^{-\gamma_k (t_{k+1} - t)} S_{t_{k+1}-t}^{-1} \Big[
  x_{k+1} - \alpha_k + e^{-\gamma_k (t_{k+1} - t)} \alpha_k
  -\frac{a}{\gamma_k^2 + \omega^2} \Big( (\gamma_k
  \sin(\omega t_{k+1}) - \omega \cos(\omega t_{k+1}))\\
  &- e^{-\gamma_k (t_{k+1} - t)} (\gamma_k
  \sin(\omega t) - \omega \cos(\omega t)) \Big) \Big]
  + C(t) S_{t-t_k}^{-1} \Big[ \alpha_k + e^{-\gamma_k (t -
    t_k)} (x_k - \alpha_k) \\
  &+ \frac{a}{\gamma_k^2 + \omega^2} \Big( (\gamma_k
  \sin(\omega t) - \omega \cos(\omega t))
  - e^{-\gamma_k (t - t_k)} (\gamma_k \sin(\omega
  t_k) - \omega \cos(\omega t_k)) \Big) \Big], \\
  g_t(x) &\approx f(x_k) + a \sin(\omega t) - \gamma_k (x - x_k)
  + D e^{-\gamma_k (t_{k+1} - t)} S_{t_{k+1} - t}^{-1} \Big[
  x_{k+1} - \alpha_k
  - e^{-\gamma_k (t_{k+1} - t)} (x - \alpha_k) \\
  &- \frac{a}{\gamma_k^2 + \omega^2} \Big( (\gamma_k \sin(\omega
  t_{k+1}) - \omega \cos(\omega t_{k+1})) - e^{-\gamma_k (t_{k+1} -
    t)} (\gamma_k \sin(\omega t) - \omega \cos(\omega t)) \Big) \Big].
\end{align*}

\end{widetext}

\section{MCMC sampler}
\label{sec:mcmc}

We briefly describe the Markov Chain Monte Carlo (MCMC) algorithm,
which generates samples from the drift function of a system of SDEs
with known diffusion. Similar to the EM algorithm in the main text,
the drift is modeled in a nonparametric way.

As before, our data will be a set of $N$ observations $\mathbf{Y} =
(y_1, \ldots, y_N)$, where $y_k = X_{k\tau}$. Since the time distance
between adjacent observations is taken to be large, we impute the
process between observations in interval $I_k = [k \, \tau; (k + 1)
\tau]$ on a fine grid of step size $\Delta = \tau/M$ for some suitable
integer $M$. The imputed path of the $k$th subinterval will be denoted
by $\mathbf{X}_k = \left\{X_{k\tau}, X_{k\tau+\Delta}, \ldots,
  X_{k\tau+M\Delta}\right\}$.

If we write the complete imputed path of length $MN$ as
\begin{multline*}
  \mathbf{X}=(y_0, X_{\Delta}, \ldots, X_{(M-1)\Delta},\ldots, y_1,
  \ldots, \\
  X_{(k-1)\tau+(M-1)\Delta}, y_k, X_{k\tau+\Delta},\ldots, y_N),
\end{multline*}
then the joint posterior distribution of the data and the drift and
diffusion function for a given set of observations is given by
\begin{displaymath}
  p(\mathbf{X}, f|\mathbf{Y}, D)\propto p_0(f) \prod_{l=1}^{NM}
  p(X^{l+1}|X^l, f, D)
\end{displaymath}
Here, the density $ p(\mathbf{X}, f|\mathbf{Y}, D)$ is approximately
normally distributed(see (\ref{eq:likelihood})) on the fine grid with
mean and variance given by (\ref{eq:dense}) and
(\ref{eq:densediffusion}), respectively. A straightforward way to
sample from this posterior is given by the following Gibbs sampler:

\begin{table}[h]
  \begin{algorithm}[H]
    \caption{Gibbs Sampler}
    \begin{algorithmic}[1]
      \State Initialize $f^{(0)}$ with the direct GP solution
      \For {$i = 1, \ldots, N$}
      \State Sample $\mathbf{X}^{(i)}\sim p(\mathbf{X}|\mathbf{Y},
      f^{(i - 1)}, D)$
      \State Sample $f^{(i)}\sim p(f|\mathbf{X}^{(i)}, \mathbf{Y})$
      \EndFor
    \end{algorithmic}
  \end{algorithm}
\end{table}

Here, the superscripts denote the iteration. The number of iterations
for a particular model is determined by the usual MCMC convergence
diagnostics, see for example \cite{Robert:2013:MSM}. Since an analytic
form for the imputed path distribution $p(\mathbf{X}|\mathbf{Y}, f,
D)$ does not exist, we have to resort to a Metropolis- Hasting (MH)
step. As proposal distribution $q$, we use the so-called
\emph{modified diffusion bridge} (MBD) of \cite{Durham:2002:NTM}.
Here, for each interval $I_k$ the density of a grid point $X_k^{j +
  1}$ from $\mathbf{X}_k$ is normally distributed, conditioned on
$X_k^j$ and the interval endpoint $ y_{k+1}$:
\begin{multline}
  q(X_k^{j + 1}|X_k^j, y_{k+1}, f_q, D_q)=\\
  \mathcal{N}(X_k^{j + 1}|X_k^j+f_q(X_k^j)\Delta, D_q(X_k^j))
  \label{eq:MDB}
\end{multline}
with drift and diffusion
\begin{displaymath}
  f_q(X_k^j)=\frac{y_{k+1}-X^j}{\tau - j\Delta},\quad
  D_q(X_k^j) = \frac{\tau - (j + 1)\Delta}{\tau - j \Delta}D(X_k^j).
\end{displaymath}
Now, since for each subinterval $I_k$ the bridge proposal starts in
observation $y_k$ and terminates in $y_{k+1}$, we can generate a
sample of the complete path $p(\mathbf{X}|\mathbf{Y}, f, D)$ by
sampling a MDB proposal separately for each the $N$ subintervals.
Specifically, for subinterval $I_k$ we simulate a path
$\mathbf{X}^*_k$ on the dense grid by recursively sampling from
(\ref{eq:MDB}) and move from current state $\mathbf{X}_k$ to
$\mathbf{X}^*_k$ with probability
\begin{multline*}
  \alpha(\mathbf{X}_k, \mathbf{X}^*_k) = \min\left\{1,
   \left[\prod_{j=1}^{M-1}\frac{p(X_k^{*(j + 1)}|X_k^{*j}, f,
       D)}{p(X_k^{j + 1}|X_k^j, f, D)}\right]\right. \\
  \left.\times \left[\prod_{j=1}^{M-2}\frac{q(X_k^{j + 1}|X_k^j,
        y_{k+1}, f_q, D_q)}{q(X_k^{*(j + 1)}|X_k^{*j}, y_{k+1}, f_q,
        D_q)}\right]\right\},
\end{multline*}
with probability $(1-\alpha(\mathbf{X}_k, \mathbf{X}^*_k))$ we retain
the current path $\mathbf{X}_k$.

The sampling from the drift $p(f|\mathbf{X}, \mathbf{Y})$ is easier to
accomplish, since under a GP prior $p_0 \sim \mathcal{GP}$ assumption,
the distribution $p(f|\mathbf{Y},\mathbf{X})$ of the SDE drift
corresponds to a GP posterior and is therefore of analytic form. Since
the number of dense path observations is usually quite substantial, we
resort to the sparse version of the GP with mean and variance given by
(\ref{eq:sparsedense}) and (\ref{eq:sparsedensediffusion}),
respectively. In each iteration of the Gibbs sampler, we simulate a
new $f$ on a fine grid over the (slightly extended) range of the path
observations $\mathbf{X}$ and then interpolate these points by
nonparametric regression in order to arrive at an approximate drift
function. The interpolation step, for which we again resort to a
sparse GP, is motivated by computational considerations, since this
way evaluating the function values for the path can be can be done
very efficiently, while also being accurate due to the smoothness of
the underlying drift.

\bibliographystyle{unsrt}
\bibliography{NonparSDE}

\end{document}